\documentclass[aps,prd,showpacs,preprintnumbers]{revtex4}
\usepackage{graphics}
\usepackage{multirow}
\usepackage{amsmath}
\newcommand{\Rmnum}[1]{\romannumeral #1\ }
\newcommand{\ie}{{\sl i.e.\/}}

\newcommand{\beq}{\begin{equation}}
\newcommand{\eeq}{\end{equation}}
\newcommand{\beqa}{\begin{eqnarray}}
\newcommand{\eeqa}{\end{eqnarray}}
\newcommand{\pldiff}{\delta P}
\newcommand{\plasma}{gluo$N_c$ plasma}
\newcommand{\re}{{\rm Real\/}\,}
\newcommand{\tr}{{\rm tr\/}\,}

\newcommand{\sun}{{\rm SU}(N_c)}

\newcommand{\ept}{\Delta/T^4}
\newcommand{\ep}{\Delta}
\newcommand{\epm}{\Delta_{\rm max\/}}
\newcommand{\latent}{\Delta\epsilon}

\newcommand{\stbo}{{\scriptscriptstyle SB}}
\newcommand{\esb}{\epsilon_\stbo}
\newcommand{\psb}{p_\stbo}
\newcommand{\ssb}{s_\stbo}
\newcommand{\simgt}{\,\rlap{\lower 7.5 pt\hbox{$\mathchar \sim$}}\raise 3 pt \hb
ox{$>$}\,}
\newcommand{\simlt}{\,\rlap{\lower 7.5 pt\hbox{$\mathchar \sim$}}\raise 3 pt \hb
ox{$<$}\,}

\def \ie{{\sl i.e.\/}}
\def \etal{{\sl et al.\/}}
\def \jhep{{\sl J.\ H.\ E.\ P.\/}}

\def \np{{\sl Nucl.\ Phys.\/}}
\def \pl{{\sl Phys.\ Lett.\/}}
\def \pr{{\sl Phys.\ Rev.\/}}
\def \prl{{\sl Phys.\ Rev.\ Lett.\/}}

\begin{document}
\title{Continuum Thermodynamics of the Gluo$N_c$ Plasma}
\author{Saumen \surname{Datta}}
\email{saumen@theory.tifr.res.in}
\affiliation{Department of Theoretical Physics, Tata Institute of Fundamental
         Research,\\ Homi Bhabha Road, Mumbai 400005, India.}
\author{Sourendu \surname{Gupta}}
\email{sgupta@tifr.res.in}
\affiliation{Department of Theoretical Physics, Tata Institute of Fundamental
         Research,\\ Homi Bhabha Road, Mumbai 400005, India.}
\begin{abstract}
  We study the thermodynamics of $\sun$ pure gauge theories for $N_c=3$,
  4 and 6. The continuum and thermodynamic limits of bulk quantities such
  as the pressure ($p$), energy density ($\epsilon$) and the entropy density
  ($s$) are taken by using several different lattice spacings and volumes.
  There is no window of temperature in which a non-trivial conformal theory
  describes bulk thermodynamics.
  We extract the latent heat of the first-order deconfinement phase transitions
  and observe good scaling with $N_c$.
  For all quantities that we measure, strong $N_c$ scaling holds, except,
  possibly, very close to the transition temperature, $T_c$; however
  we are unable to find strong evidence for scaling with the 't Hooft coupling
  in thermal quantities at the small values of $N_c$ which we study.
\preprint{TIFR/TH/10-13}
\end{abstract}
\maketitle

\section{Introduction}
\label{sec.intro}

\begin{figure}[t]
\begin{center}
\scalebox{1.0}{\includegraphics{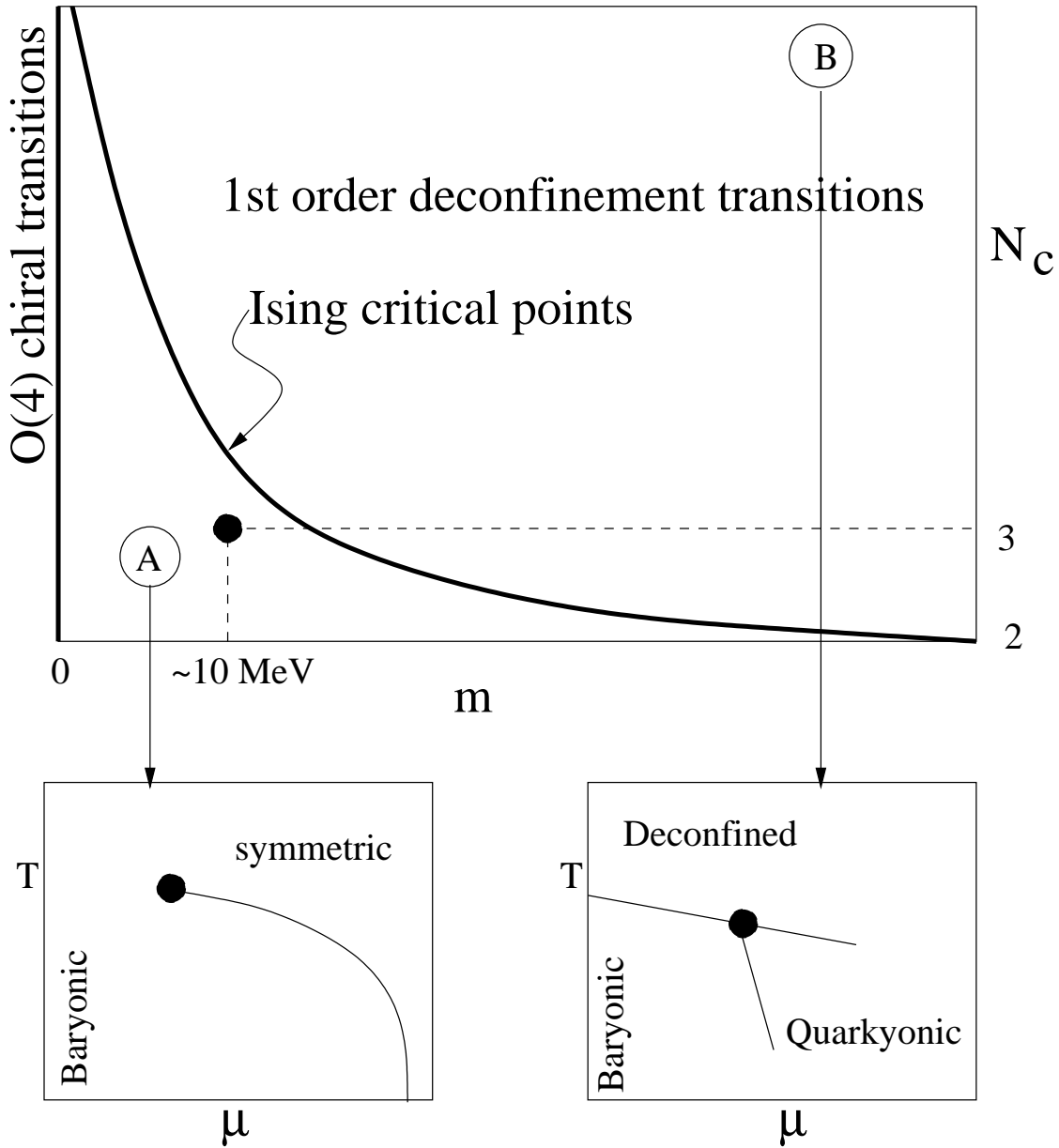}}
\end{center}
\caption{The diagram on top shows the regions of first- and second-order
  thermal phase transitions for $\sun$ gauge theories with two flavours of
  fermions as $N_c$ and the quark mass, $m$, are varied. We also show the
  phase diagrams in two interesting regions--- (A) the chiral cross over
  region and (B) the large-$N_c$ deconfinement region. The topology of the
  phase diagrams in region A is expected to involve a line of first order
  transitions ending in a critical point at finite chemical potential $\mu$.
  In region B large-$N_c$ arguments indicate that the phase diagram contains
  three lines of first order phase transitions separating pairs of phases
  labelled baryonic, deconfined and quarkyonic, with a triple point where
  all three phases coexist.}
\label{fig.flag}\end{figure}

Interest in $\sun$ gauge theories with large $N_c$ began with the
pioneering studies of \cite{thooft}, where it was shown that in $1+1$ dimensions the limit 
$N_c\to\infty$ when taken such that the gauge coupling, $g\to0$, with
the 't Hooft coupling $\lambda=N_cg^2$ fixed gave rise to an interesting
and non-trivial but tractable theory. In this so-called 't Hooft scaling limit
stable mesons exist with properties which parallel much of known hadron
phenomenology.  The limit is non-perturbative, and the computation of
any correlation function requires the summation of an infinite number of
well-characterized Feynman diagrams. Corrections to this limit appear in
powers of $1/N_c$ in general, and in powers of $1/N_c^2$ for the pure
gauge theory. Since then many models have used this 't Hooft limit
\cite{catchall}, including the currently fashionable conformal cousins
of QCD: the so-called ${\cal N}=4$ supersymmetric theories which turn out
to be tractable using the AdS/CFT correspondence. Lattice simulations of
theories with small $N_c\ge3$ can test the smoothness of approach to the
't Hooft limit.

A contemporary reason for studying the large-$N_c$ theory is in the
light it could throw on the phase diagram of QCD. At any fixed $N_c$
with two flavours of massless quarks, the theory is expected to have a
second-order chiral symmetry restoring transition at finite temperature,
$T$, which is in the O(4) universality class. If a tiny mass is given to
the quarks, then the theory has a cross-over at finite temperature instead
of a phase transition. The limit of infinite quark mass corresponds
to the pure gauge theory, which has a first-order deconfining thermal
transition for all $N_c>2$. Hence at an intermediate quark mass there
is a critical point, in the Ising universality class, which ends a line
of first-order deconfining transitions \cite{qm09}. We represent this
information in the diagram of Figure \ref{fig.flag}. In the regions
with the chiral cross-over (region A) and the large $N_c$ deconfinement
transition (region B) the phase diagrams in the plane of $T$ and baryon
chemical potential, $\mu$, are topologically distinct; a representative
phase diagram from each of these regions is also shown. In region A one
expects a first-order phase transition line dividing the chiral symmetry
broken phase from the deconfined {\plasma} phase with a critical end-point 
at finite $\mu$. In region B one might expect a triple-point
with coexistence of the three phases--- \plasma, baryonic, \ie, chiral
symmetry broken and confined, and quarkyonic, \ie, chirally symmetric
and confined \cite{largencpd}. Since the arguments for the existence of a
quarkyonic phase are based on a picture of the large-$N_c$ theory which
arises in the 't Hooft limit, it is interesting to explore the range of
validity of such large-$N_c$ arguments.

In this paper our primary interest is in examining the thermodynamics
of $\sun$ pure gauge theory in the large volume and continuum limits.
The deconfinement transition in $\sun$ theories has been studied in
\cite{wingate,gavai,teper1,teper2,scaling}.  The latent heat of the
transition was studied in \cite{gavai,teper2} on spatial volumes, $V$,
of size $T\sqrt[3]V<3$.  The equation of state (EOS) has been studied
with lattice spacing of $a=1/(5T)$  up to $2T_c$ in \cite{barak} and
up to $3.5T_c$ in \cite{panero}.  Previous work has shown that the
continuum limit of the EOS \footnote{The lattice spacing $a=1/(5T_c)$
is not far from the bulk transition for these $N_c$; the consequent
finite lattice spacing error is propagated to the pressure at all $T$
when the integral method is used.} is hard to extract on such coarse
lattices \cite{teper2,scaling}.  In view of this, we have studied
the equation of state at smaller lattice spacing in the extended
temperature range $0.97 T_c\le T\le4T_c$ (preliminary results were
presented in \cite{qm09}). We extrapolate to the continuum limit using
the non-perturbative beta-functions determined earlier \cite{scaling}.
We perform finite size scaling studies by changing the spatial volume
up to $T\sqrt[3]V=4$ for $a=1/(6T_c)$ and $T\sqrt[3]V\simeq3.5$ for
$a=1/(8T_c)$.  We use statistics significantly larger than used before
in this context, and comparable to that used in studies of thermodynamics
for SU(3) pure gauge theory.

One of the purposes of a study like this is to perform lattice simulations for
small $N_c>3$, and, from measurements of any physical quantity, find the
series expansion for it around $N_c=\infty$ in powers of $1/N_c^2$.
In this work we assume that the $N_c=\infty$ limit exists and that
there is a series expansion around it, and ask what our data imply for
the radius of convergence of this series. Technically, this also means
that for each series expansion, we ask how reliably the $N_c\to\infty$
limit can be taken from measurements for a few small values of $N_c$.

In agreement with earlier studies, we find good evidence for scaling to
$N_c\to\infty$ keeping $T/T_c$ fixed.  Such a ``strong $N_c$ scaling'',
previously observed for many quantities on the lattice, also gives small
$1/N_c$ corrections. We also examine 't Hooft's limiting procedure.  It is
clear that the notion of 't Hooft scaling has to be defined carefully in
any theory with a non-vanishing beta-function \cite{catchall}. Firstly
because one has to use a (scale dependent) renormalized coupling with its
attendant scheme ambiguities. Secondly because of this scheme dependence,
$1/N_c$ corrections may be moved between the operator expectation values
and the 't Hooft coupling. Even keeping these uncertainties in mind,
we find that scaling at fixed $\lambda$ gives large corrections (in the
expansion around $N_c=\infty$) at small $N_c$, including the physically
interesting case of $N_c=3$.

The plan of this paper is as follows. In the next section, we summarize
the formul{\ae} used for calculation of the various thermodynamic
quantities. Next, in section \ref{sec.lht} we discuss the latent heat of
the deconfinement transitions.  In section \ref{sec.e3p} we investigate
the conformal symmetry breaking measure, which is the trace of the
energy momentum tensor, $\Delta=\epsilon-3p$.  In section \ref{sec.eos}
we discuss the pressure and the remaining bulk thermodynamic quantities.
Section \ref{sec.nc} is devoted to a comparison of results from theories
with different number of colors, to get an estimate of the size of
the leading $O(1/N_c)$ corrections.  Section \ref{sec.dis} analyzes
the calculated thermodynamic quantities, to infer properties of the
\plasma.  The appendix contains a detailed discussion
of the beta-functions used in this study.

\section{Formalism and definitions}
\label{sec.detail}

The thermodynamics of the $\sun$ gauge theory can be obtained from the
partition function, 
\beq
   Z(V,T) = \int {\mathcal D}U \ {\rm exp}(-S),\qquad
   S=\beta\sum_{x,\mu,\nu<\mu} \left\{1-\re P_{\mu\nu}(x)\right\}
\label{part}\eeq
calculated on a space-time lattice with $N_s$ lattice sites in each
of the spatial directions and $N_t$ in the time direction; the lattice
sites are labelled by the 4-component index $x$ and directions by Greek
indices, $1\le\mu,\nu\le4$. The bare gauge coupling, $g^2=2N_c/\beta$
determines the lattice spacing, $a$, which is implicit in the above
equations. The spatial volume is $V=N_s^3a^3$ and the temperature is
$T=1/(aN_t)$. Since we investigate finite size effects, it is useful
to introduce the aspect ratio, $\zeta = N_s/N_t=T\sqrt[3]V$.  $P_{\mu\nu}(x)$
is the trace of the product of $\sun$ valued link matrices $U$ around
the plaquette in the $\mu,\nu$ plane starting at site $x$. The trace is
normalized such that $P_{\mu\nu}(x)=1$ if the link matrices are set to
the identity.

The expectation value of the plaquette, 
\beq
   P = \frac1{6N_s^3N_t} \sum_{x,\mu<\nu}\re P_{\mu\nu}(x),
\label{pkt}\eeq
over the ensemble at any temperature, $T$, is one of the primary
observables on the lattice. The other is the expectation value of
the Wilson line,
\beq
   L = \frac1{N_s^3}\sum_{{\mathbf x}} \tr \prod_{x_4=1}^{N_t} U_{\hat t}(x),
\label{wline}\eeq
which is the spatial average of the product of time-like link variables
wrapping around the lattice in the time direction at each spatial
site, $\mathbf x$. $\langle L\rangle$ is the order parameter of the
confinement-deconfinement transition and changes from zero to non-zero
values at the (first-order) phase transition temperature $T_c$. In the
deconfined phase $\langle L\rangle$ has $N_c$ allowed values. Often
$\langle|L|\rangle$ is examined, although it is not an order parameter,
since it has only two allowed values in the transition region: being close
to vanishing in the confined phase and non-zero in the deconfined phase.
These observables and the scaling of $T_c$ to the continuum have been
reported earlier \cite{scaling}.

The integral in (\ref{part}) is performed by Monte Carlo sampling, using
a combination of pseudo-heatbath and over-relaxation steps, where all
SU(2) subgroups of the $\sun$ group elements are touched. For details of
the algorithm used and its performance, see \cite{scaling}. We studied
SU(4) and SU(6) theories in the temperature range between $T_c$, and
about $4T_c$.  Since a major focus of this study is to get results in
the continuum and thermodynamic limit, over the whole temperature range,
we use two different lattice spacings, $a=1/(6T)$ and $1/(8T)$, at each
temperature and several different $\zeta$ \cite{epaps}.  The temperature
scale for these theories was set in \cite{scaling}, where it was shown
that near $T_c$ the results are in the scaling regime. With the running
of the coupling obtained in that study, we found very good agreement
between the thermodynamic quantities extracted on lattices with the
different $a$ at all $T$. We have also made a few simulation runs for
SU(3) gauge theories at large aspect ratios, to complement existing
studies of latent heat in SU(3) gauge theories.

Bulk thermodynamic quantities are obtained by taking suitable
derivatives of the partition function, $Z(V,T)$. In particular, the
energy density and pressure are given by 
\beq
   \epsilon = \frac{T^2}V\,\left.\frac{\partial\log Z}{\partial T}\right|_V,
   \qquad\qquad
   p = T\,\left.\frac{\partial\log Z}{\partial V}\right|_T.
\label{thermo}\eeq
The entropy density is given by the identity $s/T^3 = (\epsilon+p)/T^4$.

A quantity that is easy to calculate on the lattice is the trace of the
energy-momentum tensor, $\ep=\epsilon-3p$. This is of some interest for
models of the QCD plasma, since it is a direct measure of the conformal
symmetry breaking. Using the above relations it is easy to show that 
\beq
   \frac\Delta{T^4} = 6N_t^4\frac{\partial\beta}{\partial\log a}
      \langle\pldiff\rangle, \qquad{\rm where}\qquad \pldiff=
      P(\beta,T) - \langle P(\beta,T=0)\rangle.
\label{e3p}\eeq
The expectation value of $\pldiff$ must be taken over the finite
temperature ensemble. The subtraction of the plaquette expectation
value at $T=0$ must be done at the same lattice spacing as the finite
temperature simulation. This serves to remove an ultraviolet divergence
from the plaquette. It also makes sure that $\Delta/T^4$ vanishes at
$T=0$, since both the pressure and the energy density vanish there. The
derivative multiplying this non-perturbative factor is closely related
to the beta-function of the theory.  In the appendix we
have a discussion of the choices of beta-functions and their influence
on $\Delta/T^4$.

The suggestion of \cite{boyd} was that since $p=T\log Z/V$, in the
thermodynamic limit, the pressure can be evaluated with respect to some
reference value by integrating the plaquette expectation value---
\beq
   \frac p{T^4} = \frac1{VT^3} {\rm log} Z = \frac{p_0}{T^4}
       + \frac TV \int_{\beta_0}^\beta
   d\beta^\prime\frac{\partial\log Z}{\partial\beta}
   \simeq 6 \int_{\beta_0}^\beta d\beta^\prime \pldiff(\beta').
\label{integral}\eeq
Here $p_0$ is the pressure at the reference temperature $T_0=1/N_t
a(\beta_0)$. For the {\plasma} the pressure is expected to be very
small below $T_c$, even as close to $T_c$ as $T\sim0.9T_c$. Conventionally
one takes the reference temperature to be such a value and sets $p_0=0$,
as in the second expression, where the derivative has also been written
out explicitly. Once $\Delta$ and $p$ are known, $\epsilon$ and $s$
can be evaluated.

Asymptotic freedom in $\sun$ gauge theories lead us to expect that
at sufficiently high temperatures one should obtain a free gas of
gluons. Then
thermodynamic quantities reach their ideal gas (\ie, Stefan-Boltzmann: SB)
limits. There are lattice corrections to this limit \cite{engels}. 
When the pressure is evaluated through the integral method one has---
\beq
   \frac{\esb}{T^4} = 3\,\frac{\psb}{T^4}
        = \frac{\pi^2d_A}{15}G(N_t) \qquad{\rm where}\qquad
   G(N_t) = 1 + \frac{8\pi^2}{21} \;\frac1{N_t^2} + \cdots
\label{lsb}\eeq
and $d_A=N_c^2-1$.  Since $aT=1/N_t$, the correction terms come in
powers of the lattice spacing $a$, and vanish in the continuum limit.
The full $G(N_t)$ is also known exactly from numerical computations,
and listed in \cite{engels}. When we discuss the numerical computations
later we will use this full computation of $G(N_t)$ and not the series
expansion above. The difference between the two is about 2\% for $N_t=6$.
We draw attention to the factor of $d_A$: in the large $N_c$ limit it
is often replaced by $N_c^2$, but at the small values of $N_c$ we used,
the difference is statistically significant. We use $d_A$ in this work,
and thereby subsume some of the formally sub-leading $1/N_c$ corrections
into this scaling.

One of the pieces of physics we are interested in is the latent heat of
the transition. In a thermodynamically large volume this is defined by
the formula
\beq
   \frac\latent{T_c^4} = \displaystyle\lim_{\delta T\to0}
       \left(\frac{\epsilon(T_c+\delta T)}{T_c^4} 
            - \frac{\epsilon(T_c-\delta T)}{T_c^4}\right)
   = \displaystyle\lim_{\delta T\to0}
       \left(\frac{\Delta(T_c+\delta T)}{T_c^4} 
            - \frac{\Delta(T_c-\delta T)}{T_c^4}\right)
\label{latent}\eeq
where the second equality follows from the fact that $p$ is continuous
across a first-order phase transition.  These formul{\ae} cannot be
directly used at finite volume. We describe a method for determining
$\latent/T_c^4$ in section \ref{sec.lht}.

Some remarks about the computation are best placed here. It was
observed earlier that at lattice spacing of $a\le1/(6T_c)$ the two-loop
beta-function with a $1/N_t^2$ correction provides a good description
of the change of measured length scales with the gauge coupling $g^2$.
Therefore, one should be able to perform continuum extrapolation of
thermodynamic quantities for $T\ge T_c$ using data on lattices with
$N_t\ge6$. This expectation should be correct except if there are large
corrections in powers of $a$ to the operators involved in defining
components of the energy-momentum tensor. We show evidence later that
there are no such large corrections. Another possible subtlety could
reside in having to take the $V\to\infty$ (thermodynamic) limit before
taking the continuum ($a\to0$) limit. Such subtleties arise only when
there are large correlation lengths, $\xi$. Here we have a first order
phase transition with $\xi\le1/T$ \cite{a1pp,olaf}. Hence the full
machinery of finite-size scaling need not be invoked if due caution is
exercised: the extraction of the latent heat is one such case. We will
mention these tests at appropriate places in the remainder of the paper.

\section{The latent heat}
\label{sec.lht}

\begin{figure}[thpb]
\begin{center}
\scalebox{0.6}{\includegraphics{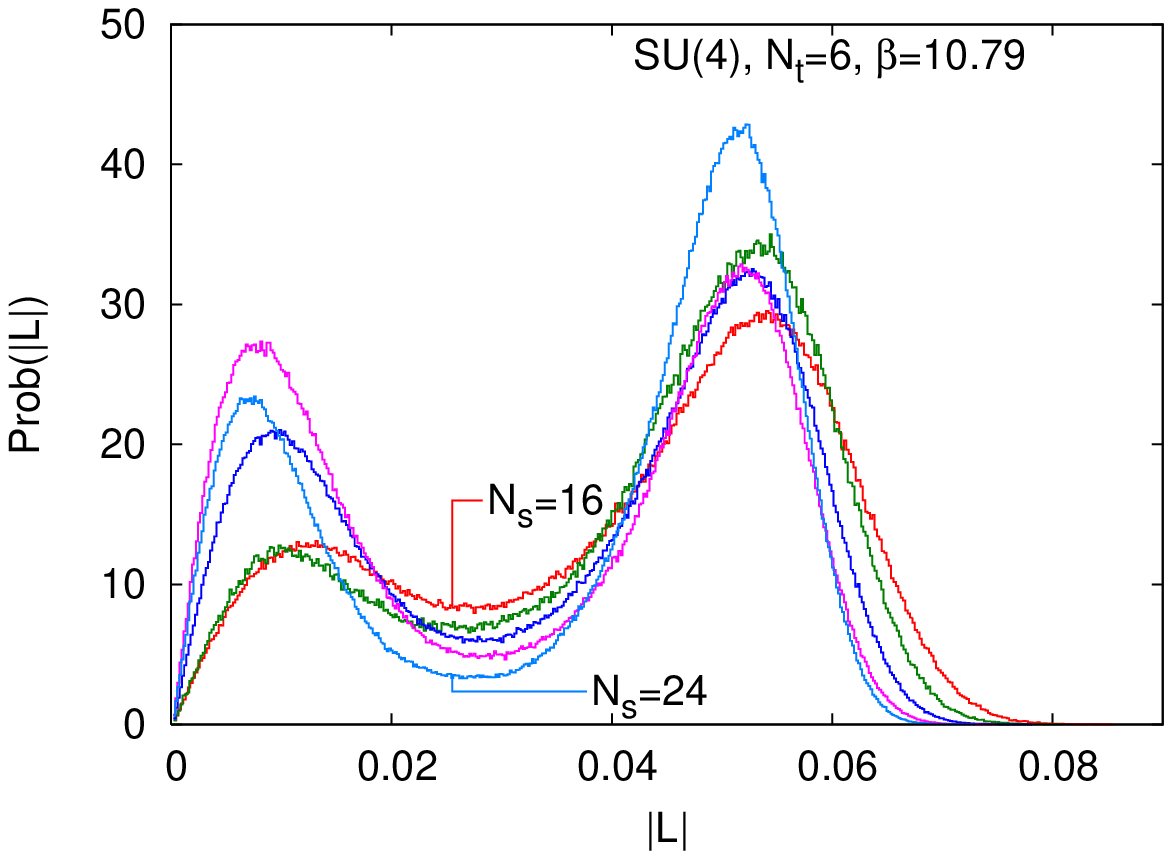}}
\scalebox{0.6}{\includegraphics{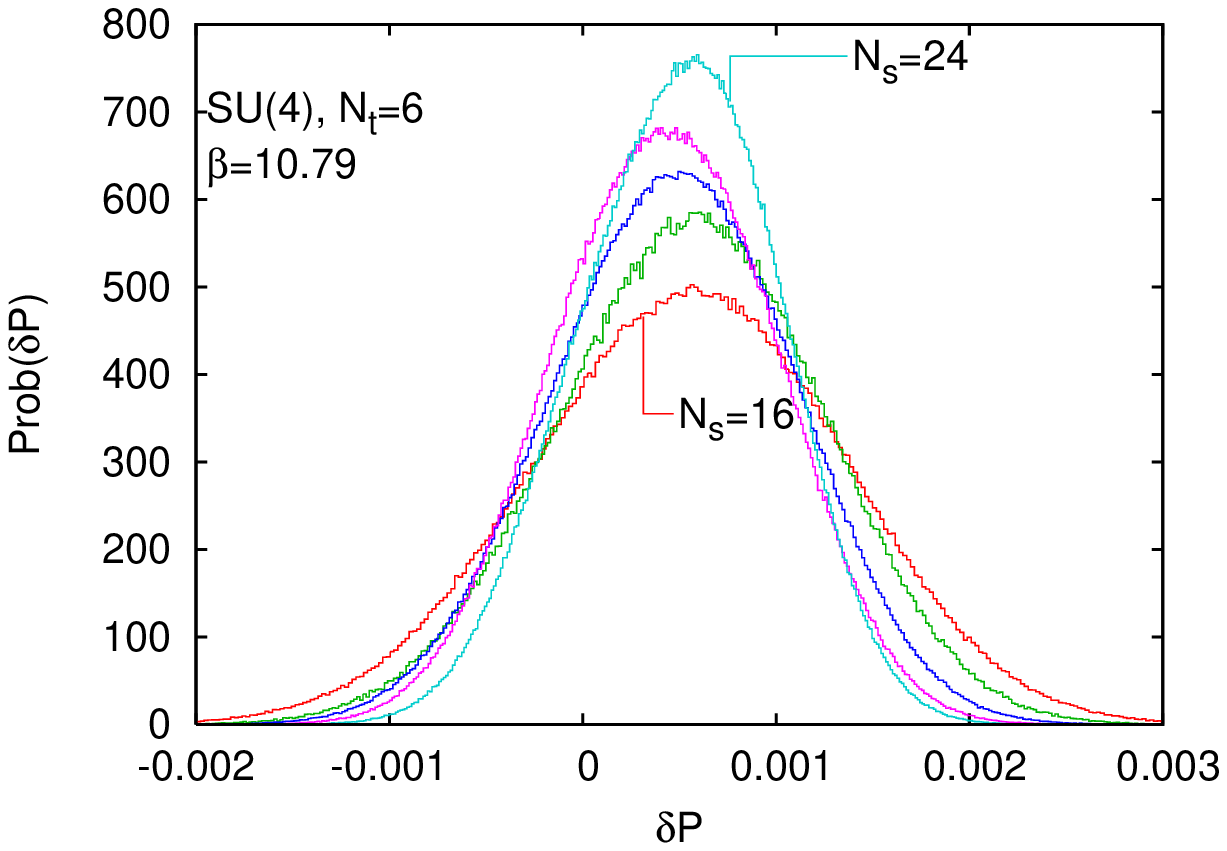}}
\end{center}
\caption{Histograms of $|L|$ and $\pldiff$ normalized to have unit
  area under the curves. A multi-peak structure in the histogram
  corresponds to multiple (local) minima of the constrained free
  energy. As $N_s$ changes between the values shown, in steps of 2,
  the histograms move up or down monotonically in the region where
  they are labelled.}
\label{fig.histo}\end{figure}

\begin{figure}[bhpt]
\begin{center}
\scalebox{0.6}{\includegraphics{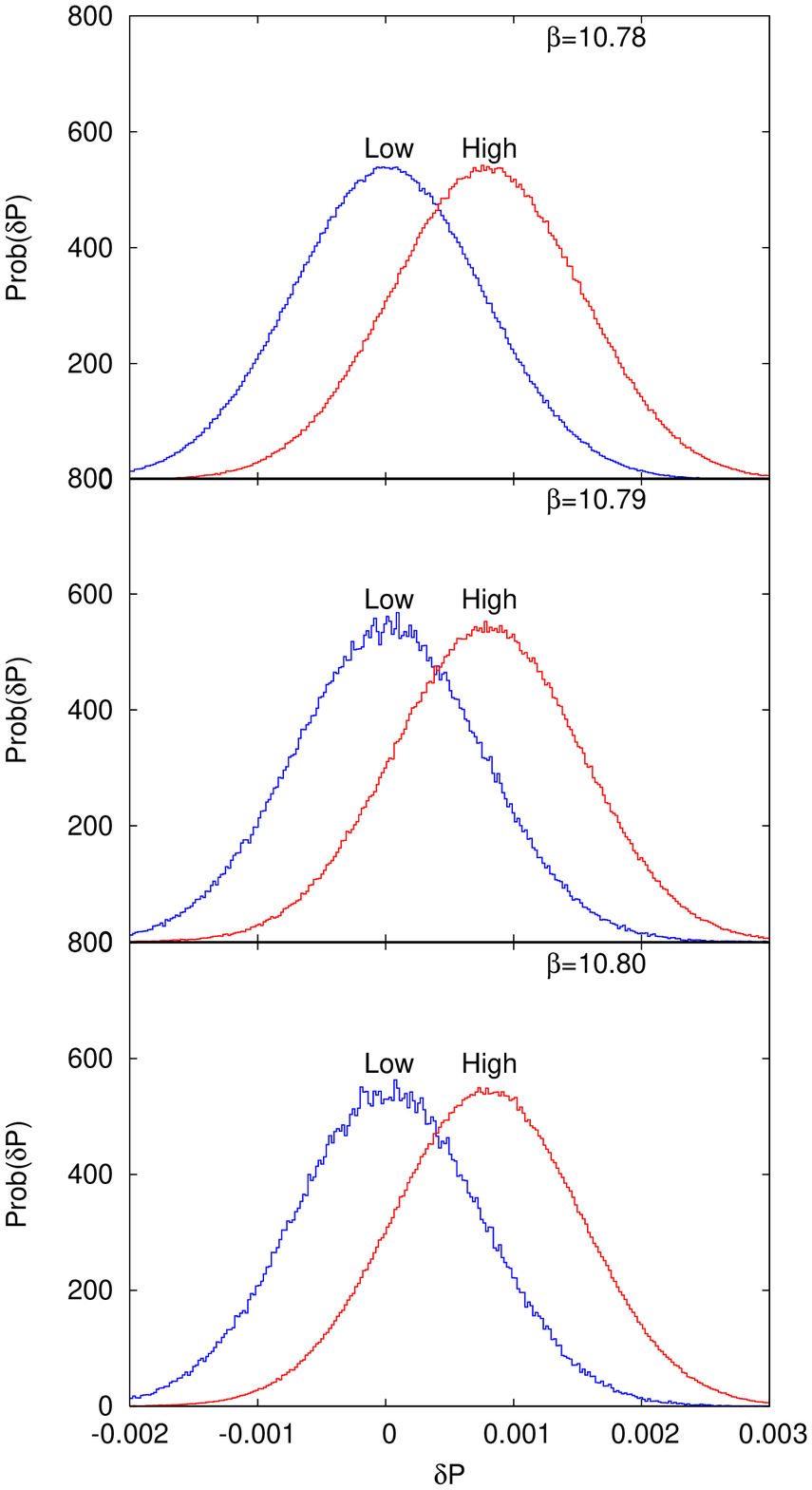}}
\scalebox{0.6}{\includegraphics{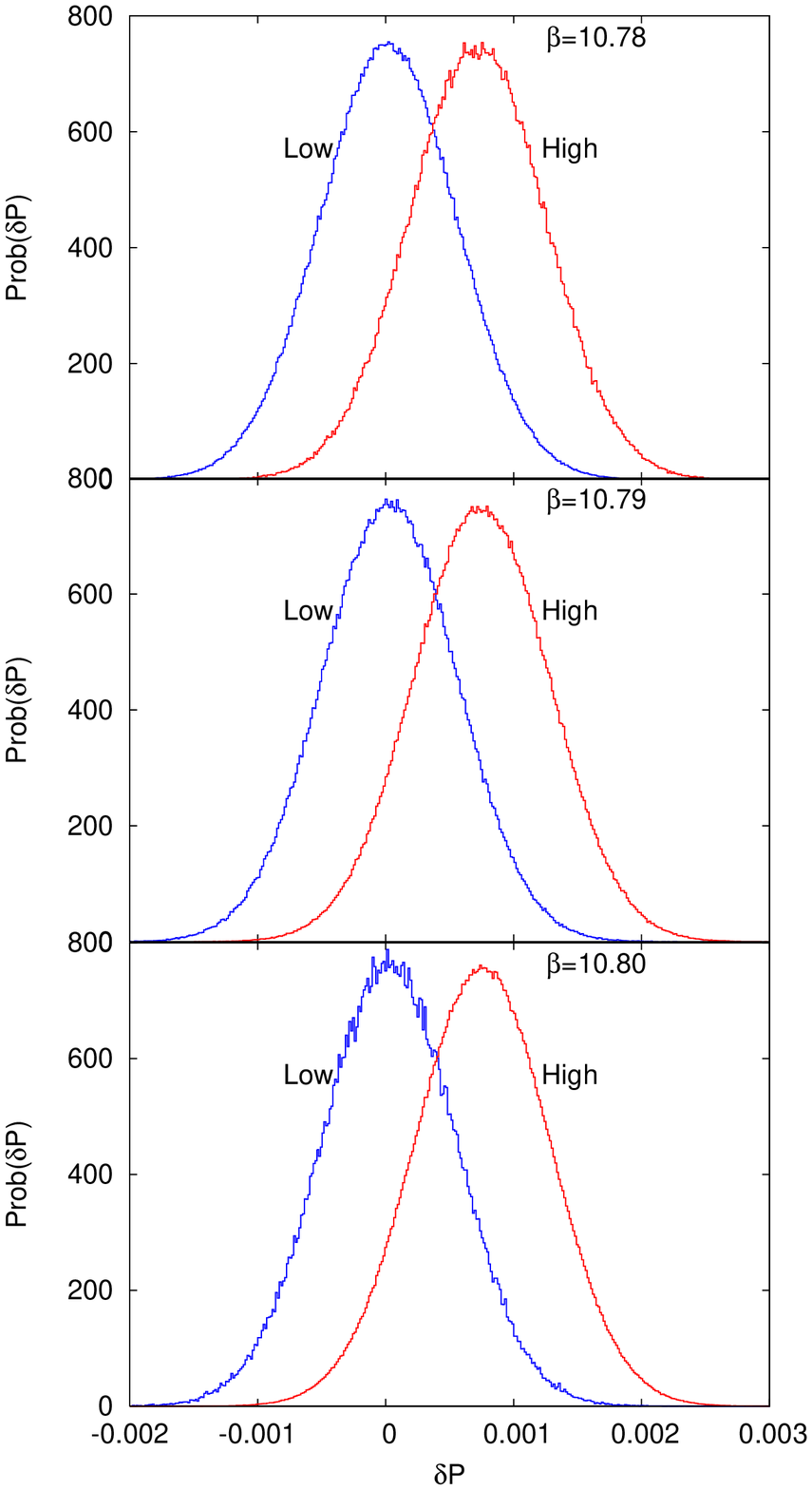}}
\end{center}
\caption{Histograms of $\pldiff$ in the two phases of SU(4) gauge theory
  as defined using $L_c=0.024$ and $L_h=0.031$. The panels on the left
  are results from $6\times16^3$ lattices and those on the right from
  $6\times20^3$. The histograms are normalized so that the area under
  each is unity. The pure phase histograms at three different couplings
  exactly overlap on each lattice.}
\label{fig.su4lht}\end{figure}

When two copies of a system with a first-order transition are held at
temperatures $T_c\pm\delta T$, the difference in their energy densities is
finite. Since the transition is of first-order, this difference remains
finite even when $\delta T\to0$, provided one has thermodynamically
large systems. When $V$ is large but not infinite, the correct limit
may be obtained as long as $\delta T/T_c$ is much larger than ${\cal
O}(\xi^3/V)$. If $V$ is large enough, then this allows one to come close
enough to $T_c$ to have confidence in the result.  For smaller $\xi^3/V$,
this procedure breaks down. By examining the reason for the breakdown,
we can develop a different procedure to estimate the latent heat.

Finite-size effects at first-order phase transitions are best understood
in terms of constrained free energies, $F_c$. This is the free energy of a
system in which $|L|$ or $\pldiff$ are restricted to have fixed values but
other parameters are allowed to vary according to the temperature. $F_c$
near a first-order phase transition has multiple minima: the deepest
corresponds to the value the constrained variable has in the stable phase
when $V\to\infty$. As $T$ changes, the depth of the minima change, and
on crossing $T_c$ the deepest minimum flips. However, because the finite
system sees finite barriers between the minima, the system explores all
the phases. Hence the discontinuity is rounded off. As the thermodynamic
limit, $V\to\infty$, is approached, the ``wrong phase'' minimum becomes
infinitely higher and the barrier separating it from the true vacuum
also becomes infinitely high. As a result, the transition sharpens and
gives the correct thermodynamic limit.

The histogram of an observable obtained from its Monte Carlo history is
proportional to $\exp(-VF_c/T)$. The object of a finite-size scaling
study of something like the latent heat is to be able to identify
the thermodynamically stable phase from histograms such as those in
Figure \ref{fig.histo}.  The same figure also illustrates the problem
which is usually faced in computing the latent heat in $\sun$ gauge
theories. Although the histogram of $|L|$ has multiple
well-identifiable maxima, the histogram of $\pldiff$ has a single
peak. If the specific heat in each of the phases is large, then the
two-peak nature of $\Delta/T^4$ could well be hidden until extremely
large volumes are reached.  Finite-size scaling methods were developed in
the past to extract reliably the specific heat when multiple maxima are
clearly developed \cite{berg,ohlsson}. However they are not applicable here,
and we need to use different techniques. We adapt one which was first
applied to gauge theories in \cite{fukugita}.

Since the phases are well resolved by $|L|$, we can try to use the
following criterion. The cold confined phase could be identified by
requiring that $|L|\le L_c$, and the hot deconfined phase by $|L|\ge
L_h$. The results of such a phase separation are stable as long as $L_c$
and $L_h$ both lie in the valley between the peaks of $|L|$.
Examples of the probability density of $\pldiff$ obtained in the
cold and hot phases so defined are shown in Figure \ref{fig.su4lht}.
The figures illustrate the fact that these probability densities are very
stable--- at each fixed volume the pure phase probability distributions
of $\pldiff$ are identical for a range of couplings around $T_c$. If we
had not separated out the panels for different $\beta$, the curves for
the three different cases would have been indistinguishable apart from
small statistical fluctuations. Furthermore,
as one changes $V$ (with fixed $a$) the mean value of $\pldiff$ in the
hot and cold phases remain the same. It appears that the difference
between the values of $\pldiff$ in the two phases is very stable under
the variation of both $T$ and $V$ near $T_c$.  It also turns out to be
fairly stable under changes of $L_c$ and $L_h$.

The agreement of the histograms of $\pldiff$ for the different $\beta$
in the transition regime show that it is possible to reliably extract
the limiting values of $\Delta/T^4$ in each of the phases. Examination of
(\ref{e3p}) shows that knowing the difference in $\pldiff$ between the hot
and cold phases one can extract easily the jump in $\Delta/T^4$ and hence
the latent heat density, since the remaining factors are well-understood.
In Table \ref{tbl.lht} we summarize our results for the latent heat for
$\sun$ gauge theories with $N_c=3$, 4 and 6.  In the entry for the latent
heat, the first error is statistical while the second is a systematic
error, \ie, the change in the result if $L_h-L_c$ changes by $\pm$ 20\%.
The results of \cite{gavai} are higher, but those of \cite{teper2} are
consistent with ours, within the larger statistical and systematic errors
of that study.

\begin{table}[htb]
\begin{center}
\begin{tabular}{lllllllll}
\hline
$N_c$ & $N_t$ & $N_s$ & $\beta$ & $|L|_c$ & $|L|_h$ &
$\latent/T_c^4$ & $\latent/\epm$ \\
\hline
\multirow{7}{*}{3}  & \multirow{3}{*}{4} & 16 & 5.6908 & 0.055 & 0.075 & 2.06(1)(3) & \\
  & & 24 & 5.6919 & 0.055 & 0.075 & 1.93(1)(3) & \\
  & & 32 & 5.6922 & 0.055 & 0.075 & 1.90(2)(2) & \\
\cline{2-8} 
  & \multirow{3}{*}{6} & 24 & 5.8934 & 0.02 & 0.03 & 1.79(2)(4) & 0.65(2) \\
  & & 32 & 5.8938 & 0.02 & 0.03 & 1.54(2)(5) & \\
  &  & 48 & 5.8940 & 0.022 & 0.032 & 1.44(4)(3) & \\
\cline{2-8} 
  & 8 & 32 & 6.0609 & 0.013 & 0.019 & 1.67(4)(4) & 0.68(3) \\
\hline
\multirow{6}{*}{4}  & \multirow{3}{*}{6} & 16 & 10.79 & 0.024 & 0.031 & 4.85(5)(6) & 0.88(2) \\
  & & 20 & 10.79 & 0.024 & 0.031 & 4.64(4)(5) & \\
 &  & 24 & 10.79 & 0.024 & 0.031 & 4.57(4)(3) & 0.85(2) \\
\cline{2-8} 
  & \multirow{3}{*}{8} & 22 & 11.08 & 0.013 & 0.017 & 4.58(5)(6) & \\
  & & 24 & 11.08 & 0.013 & 0.017 & 4.32(6)(6) & 0.82(2) \\
  & & 28 & 11.08 & 0.013 & 0.017 & 4.33(8)(6)) & \\
\hline
  & \multirow{2}{*}{6}& 14 & 24.84 & 0.025 & 0.03 & 12.20(10)(4) & \\ 
6  & & 18 & 24.84 & 0.025 & 0.03 & 12.47(4)(2) & 0.92(2) \\ 
\cline{2-8} 
  & 8 & 20 & 25.46 & 0.012 & 0.015 & 11.93(34)(5) & 0.90(3) \\  
\hline
\end{tabular}
\end{center}
\caption{The latent heat of $\sun$ gauge theories for $N_c=3$, 4 and 6.
  The thermodynamic limit of $\latent/T_c^4$ is seen to be under control
  for $N_c=4$ and 6, as is the continuum limit. For all $N_c$, the ratio
  $\latent/\epm$ scales well. The numbers in brackets are the errors on the
  least significant digits, the first is statistical and the second
  systematic.}
\label{tbl.lht}\end{table}

In SU(4) gauge theory we found that the extracted value of $\latent/T_c^4$
is fairly stable at fixed $a$ as we change $V$. In fact, as is clear
from Table \ref{tbl.lht}, when $a=1/(6T_c)$ we found that there is no
statistically significant change in the estimate of this quantity for
$\zeta\ge3$. The results for $a=1/(8T_c)$ are consistent with this
conclusion.  For the SU(3) theory, on the other hand, it seems that
$\zeta\ge6$ is needed for an estimate of the latent heat with equally
small systematic errors, \ie, finite volume corrections are larger for
the SU(3) theory. One sees that in going from $N_c=3$ to 4 the latent
heat density scales faster than $d_A$. This ties in with the intuition
developed from a study of correlation lengths \cite{a1pp,olaf} that
the SU(3) theory is weakly first-order. One expects that correlation
lengths should also become shorter with increasing $N_c$ \cite{teper2},
and hence finite volume effects should be less pronounced.

Following the analysis of the appendix, we understand that the lack
of clear scaling of $\latent/(d_AT^4)$ to the continuum limit is
not due to the use of an inappropriate beta-function. One possible
explanation is that the large finite volume effects mask the approach to
the continuum. If so, then one should be able to eliminate it by using
another quantity with the same effect. Since $\Delta/T^4$ has a very sharp
peak as a function of $T$, and that measurement would also be related to
the latent heat, we list the quantity $\latent/\epm$ in the table. As
one can see, this ratio has much better scaling properties, and the
thermodynamic and continuum limits of the ratio are very well determined.

The data collected
for $N_t=8$ in Table \ref{tbl.lht} is fitted extremely well by the form
\beq
   \frac{\latent}{d_AT_c^4} = 0.388 (3) - \frac{1.61 (4)}{N_c^2},
\label{fitted}\eeq
where the numbers in brackets are the statistical errors on the
last digit of the central values.  Interestingly, the fit yields
$\latent/(d_AT_c^4)=0.014\pm0.014$ at $N_c=2$, where there is a second
order finite temperature transition, and hence $\latent=0$.  If one
adds a term of ${\cal O}(1/N_c^4)$, then the fit changes marginally:
the limiting value for $N_c\to\infty$ is stable at the $3\sigma$
level. The coefficient of the ${\cal O}(1/N_c^2)$ term changes by 16\%,
and the next correction term is marginal, its value being less than 10\%
of the total for $N_c=3$. The extended series extrapolated to $N_c=2$
is still consistent with vanishing latent heat of this theory.  Although
$N_c=2$ must be the limit of the validity of the series expansion around
$1/N_c=0$, it seems to be well-behaved at $N_c=3$.  In agreement with
this, a reliable value of the $N_c\to\infty$ limit of $\latent/(d_AT^4)$
can be extracted.  This is an example of successful strong scaling;
$\latent/(d_AT_c^4)$ is well described by just two terms of the series
in $1/N_c^2$ even at small $N_c$.

The series for the fourth root of the above quantity may be
of interest, since $\sqrt[4]\latent$ has mass dimension unity
\cite{teper1,teper2}. This gives
\beq
   \left(\frac{\latent}{d_AT_c^4}\right)^{1/4} =
        0.798 (7) - \frac{1.09 (9)}{N_c^2}.
\label{fittedprm}\eeq
The quality of the fit, as judged by the value of $\chi^2/$DOF, is worse,
but still within the limits of acceptability.  This result above seems
to have improved convergence properties around $N_c=\infty$. However, on
extending the fit to include the ${\cal O}(1/N_c^4)$ term, we find that
the correction terms are unstable against changes. The coefficient of the
second term reduces to half its value, and the coefficient of the third
term is 6--7 times larger. The values of these terms are comparable to
each other for $N_c=3$, opening the possibility that even higher order
terms, or a resummation of the whole series, need to be taken into
account. From the previous analysis it seems that the series expansion
for $\latent/(d_AT_c^4)$ comes close to performing this resummation.
We shall show later that other mass dimension four quantities such as
$p$, $\epsilon$ and $s$ also have good strong scaling properties.

\section{Conformal symmetry breaking}
\label{sec.e3p}

\begin{figure}[tphb]
\begin{center}
\scalebox{0.6}{\includegraphics{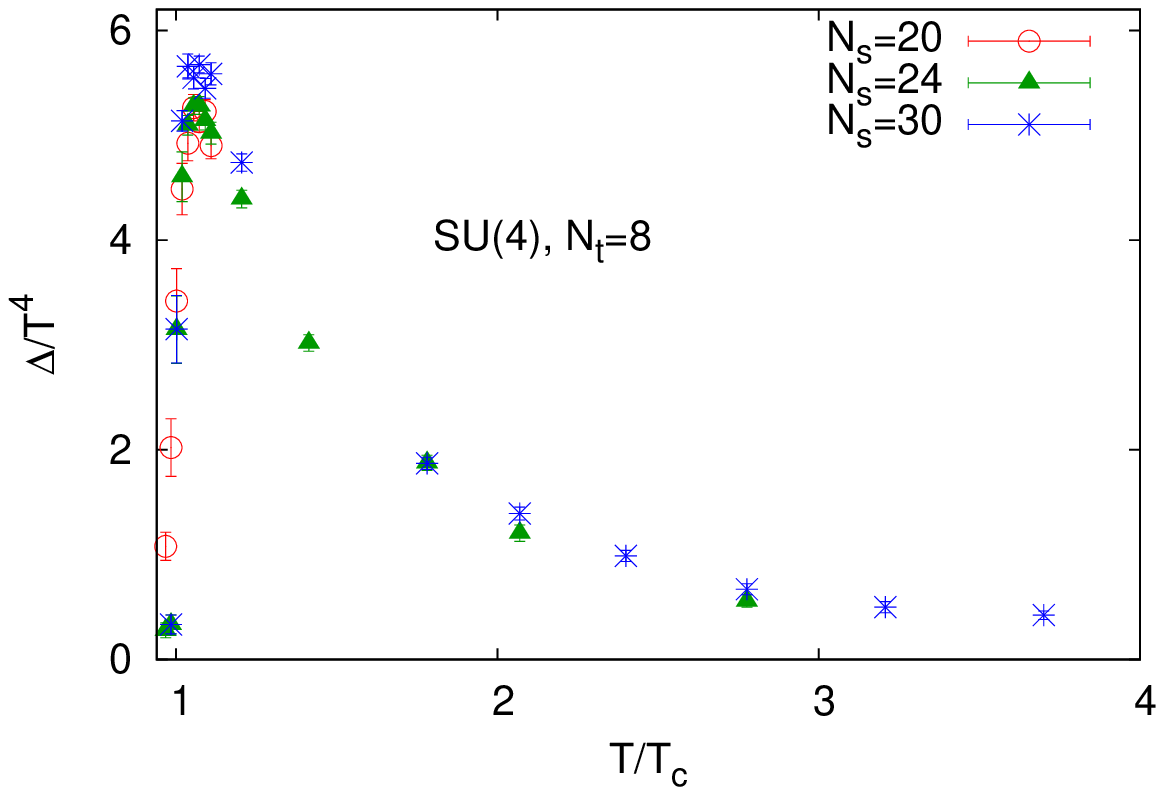}}
\scalebox{0.6}{\includegraphics{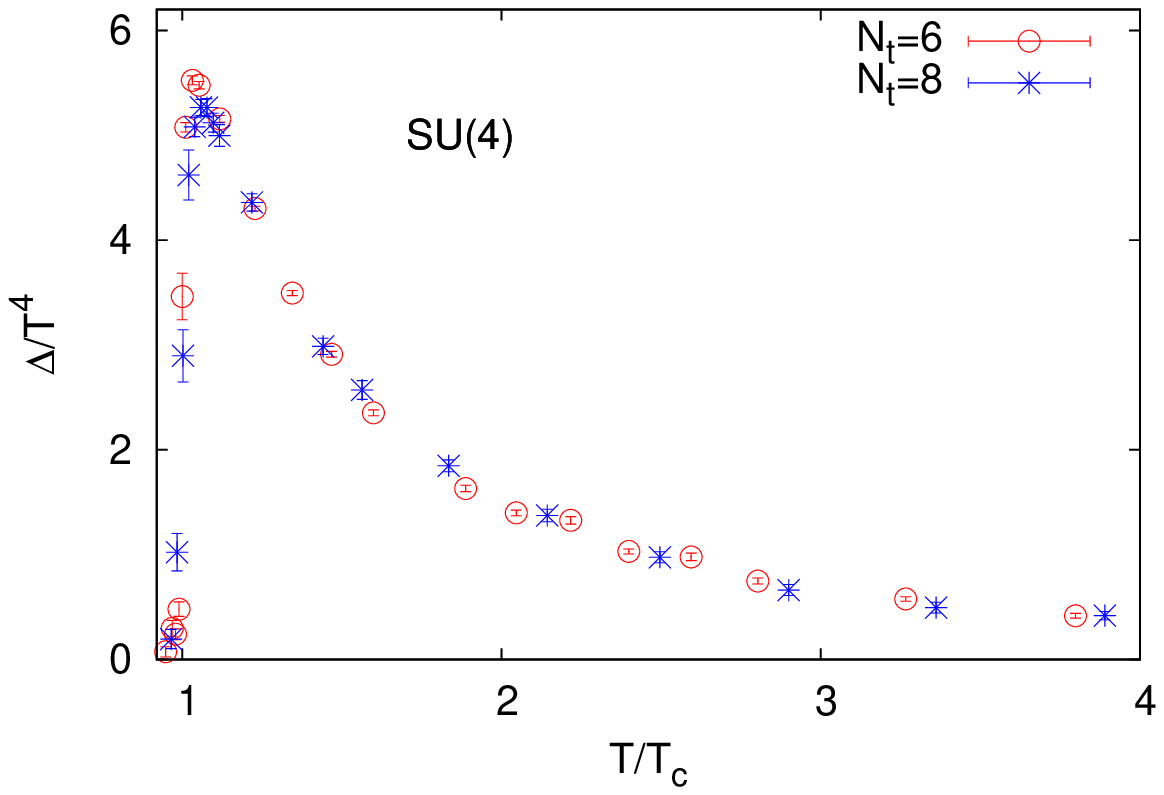}}
\scalebox{0.6}{\includegraphics{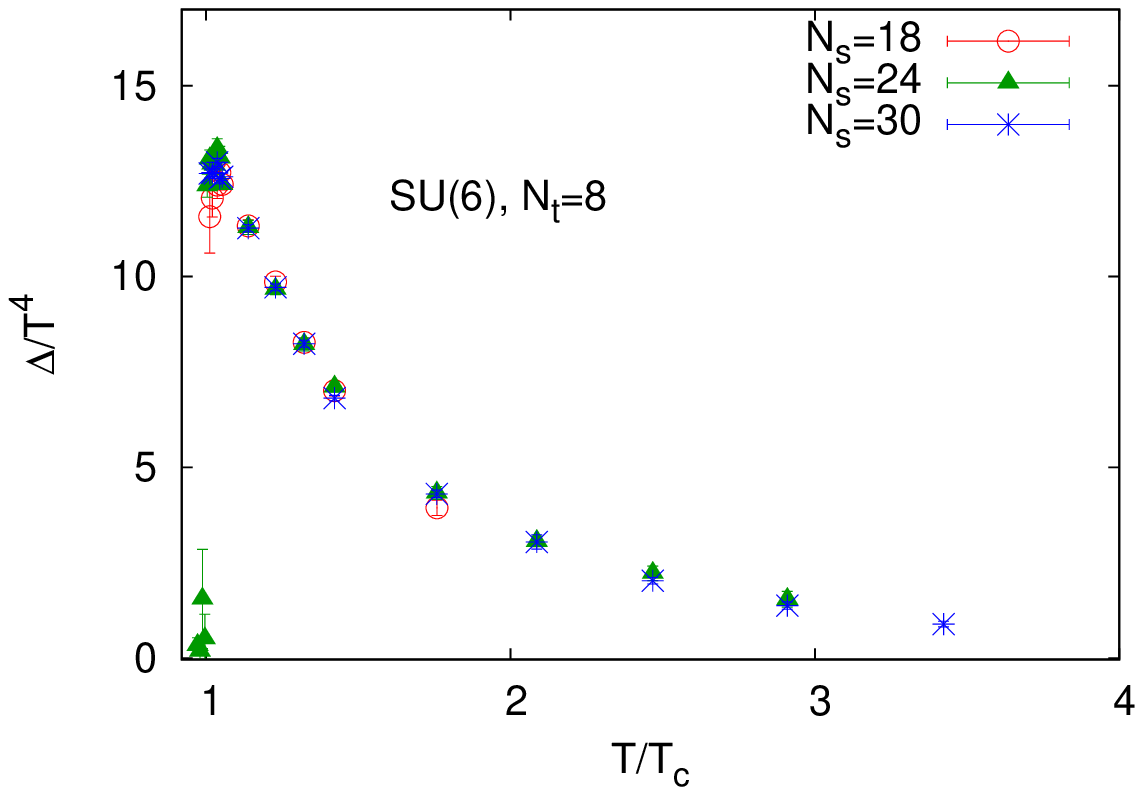}}
\scalebox{0.6}{\includegraphics{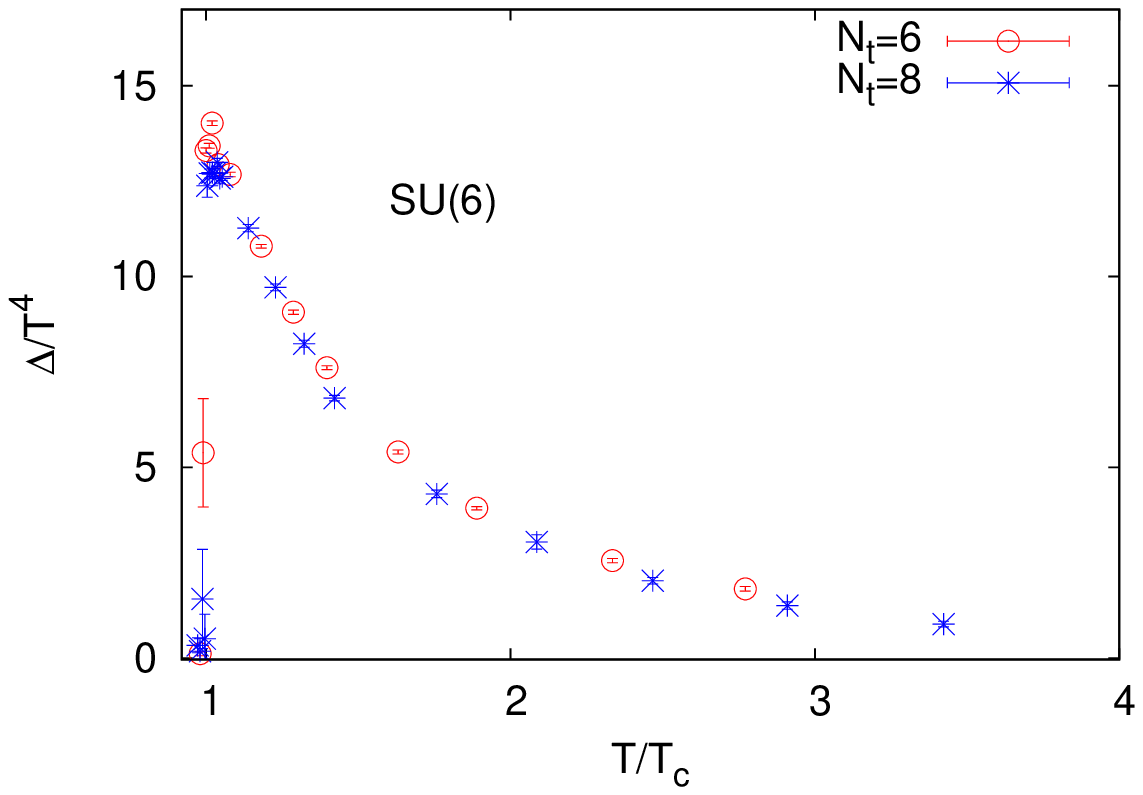}}
\end{center}
\caption{Systematics of $\Delta/T^4$ in SU(4) and SU(6) lattice gauge
  theories. Both finite size and finite lattice spacing effects are small, 
  except, possibly, very close to $T_c$.}
\label{fig.e3p}
\end{figure}

\begin{figure}[bhpt]
\begin{center}
\scalebox{0.9}{\includegraphics{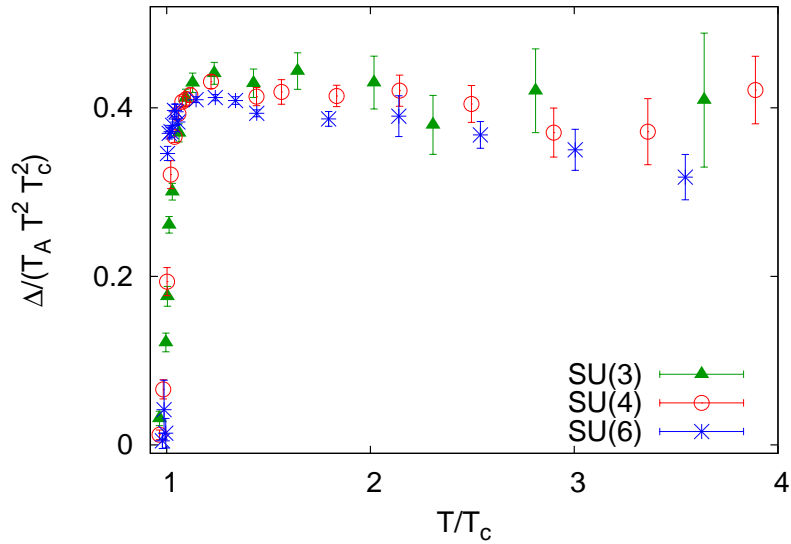}}
\end{center}
\caption{Testing a different scaling \cite{pisarski} of the conformal
  symmetry breaking measure: $\Delta/(d_AT^2T_c^2)$, where $d_A=N_c^2-1$,
  for $N_c=3$, 4, and 6.}
\label{fig.suNrob}\end{figure}

As discussed in section \ref{sec.detail}, $\ept$ is easily calculated
on the lattice.  We have seen in the case of the latent heat, however,
finite volume and cutoff effects need to be
controlled.  In the SU(3) gauge theory, it is known that $\ept$ 
rises rapidly near $T_c$ and peaks at about $1.1 T_c$ \cite{boyd}. The
rapid rise is, of course, dictated by the existence of a latent heat,
but the shift in the peak away from $T_c$ is not yet understood. We
examine the $N_c$ dependence of this peak.

In Figure \ref{fig.e3p}, we show the sensitivity of $\ept$ to $\zeta$ and
$a$. Except in the immediate vicinity of $T_c$ there seems to be little
sensitivity to $\zeta$. The cutoff dependence is also insignificant,
except in the vicinity of the peak, $\epm$.  The good agreement between
data obtained with $a=1/(6T)$ and $1/(8T)$ show that the results of the
measurement with either lattice spacing can be taken to be an estimate of
the continuum limit. We choose the conservative alternative of using $a=1/(8T)$
as a determination of the continuum results. The peak of $\Delta/T^4$
for SU(4) gauge theory is in the range $1.04\le T/T_c\le1.08$.

The systematics of $\ept$ for SU(6) gauge theories is also shown in the
same figure. The trends are very similar to those in SU(4). Results for
different $\zeta$ agree very well. The approach to the continuum limit
is also very similar to that discussed for SU(4). Again, in this case,
we can take the results obtained with $a=1/(8T)$ to be an estimate of
the continuum limit. In going from $N_c=3$ to 6, the peak in $\Delta/T^4$
moves closer to $T_c$.

It is of phenomenological interest to note that $\Delta/T^4$ is not small
even at $2T_c$. In fact, as one can see in Figure \ref{fig.e3p}, one has
\beq
   \left.\frac{\ep}{T^4}\right|_{2T_c} \simeq 0.1 d_A.
\label{splvalue}\eeq
For $N_c=3$ this implies $\ep^{1/4}\simeq T=2T_c$. This is a natural scale,
and therefore the theory is far from conformal.

We end this section with the investigation of an intriguing observation
made in \cite{ogilvie,pisarski}: in the temperature range $1.1T_c\le
T \le 4T_c$, for the SU(3) pure glue plasma, $\Delta/T^2$ seems to
be roughly constant. Phenomenological models of the gluon plasma
have introduced mass scales and obtained such a behaviour \cite{pheno}.
We investigated this modified scaling behaviour
at larger $N_c$ (see Figure \ref{fig.suNrob}).  The dimensionless
quantity $\Delta/(d_AT^2T_c^2)$ for SU(3) gauge theory is seen to
have little temperature dependence from just above $T_c$ to about
$4T_c$. Unfortunately, the data for SU(3) theory is noisy at larger $T$.
For SU(6) the error bars are smaller and one can observe that this
quantity falls with $T$. This implies that the temperature dependence
of $\ep$ could be slower than $T^2$. It would be interesting in future
to expand the range of $N_c$ and $T$ in order to study this further.

\section{Other bulk thermodynamic quantities}
\label{sec.eos}

\begin{figure}[thpb]
\begin{center}
\scalebox{0.6}{\includegraphics{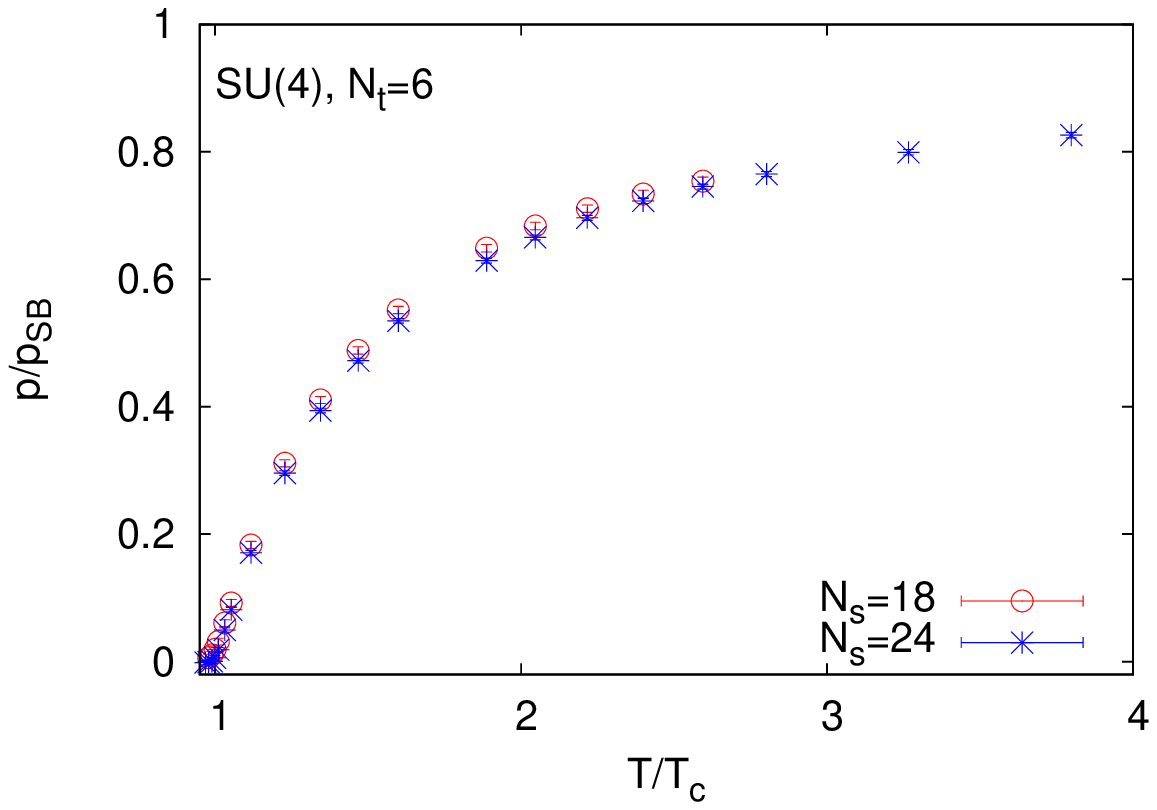}}
\scalebox{0.6}{\includegraphics{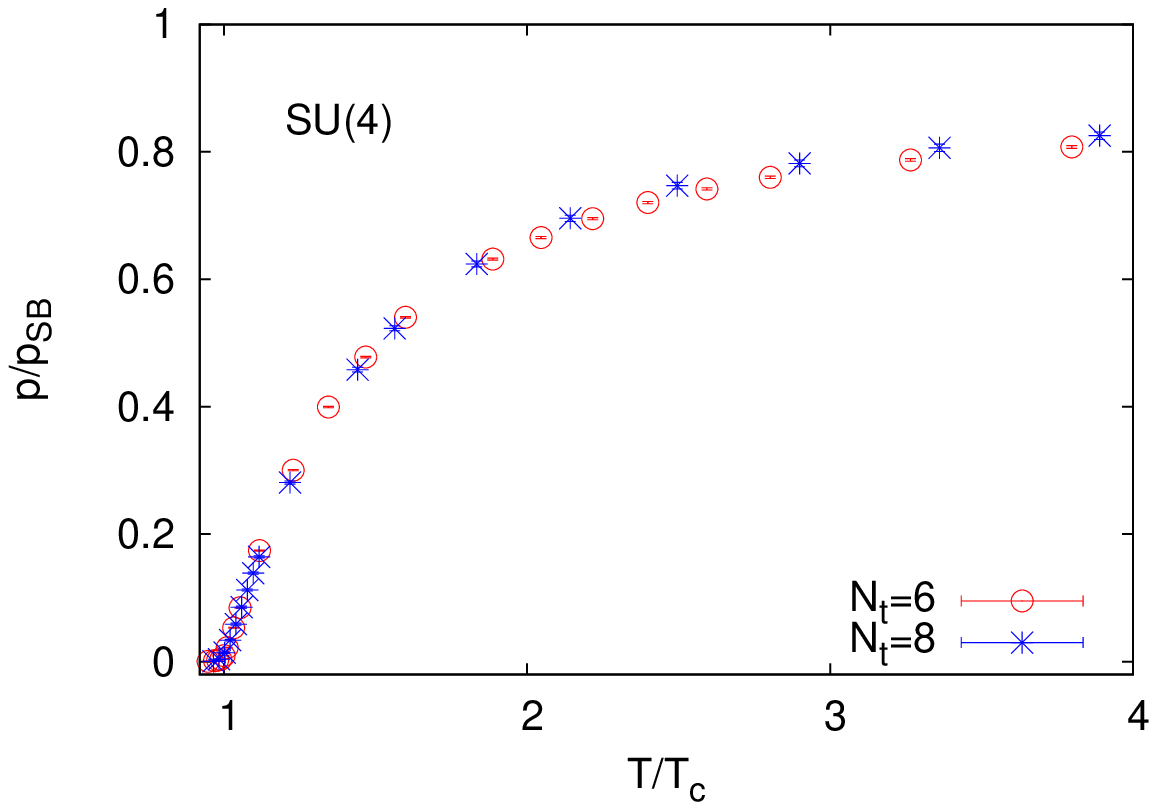}}
\scalebox{0.6}{\includegraphics{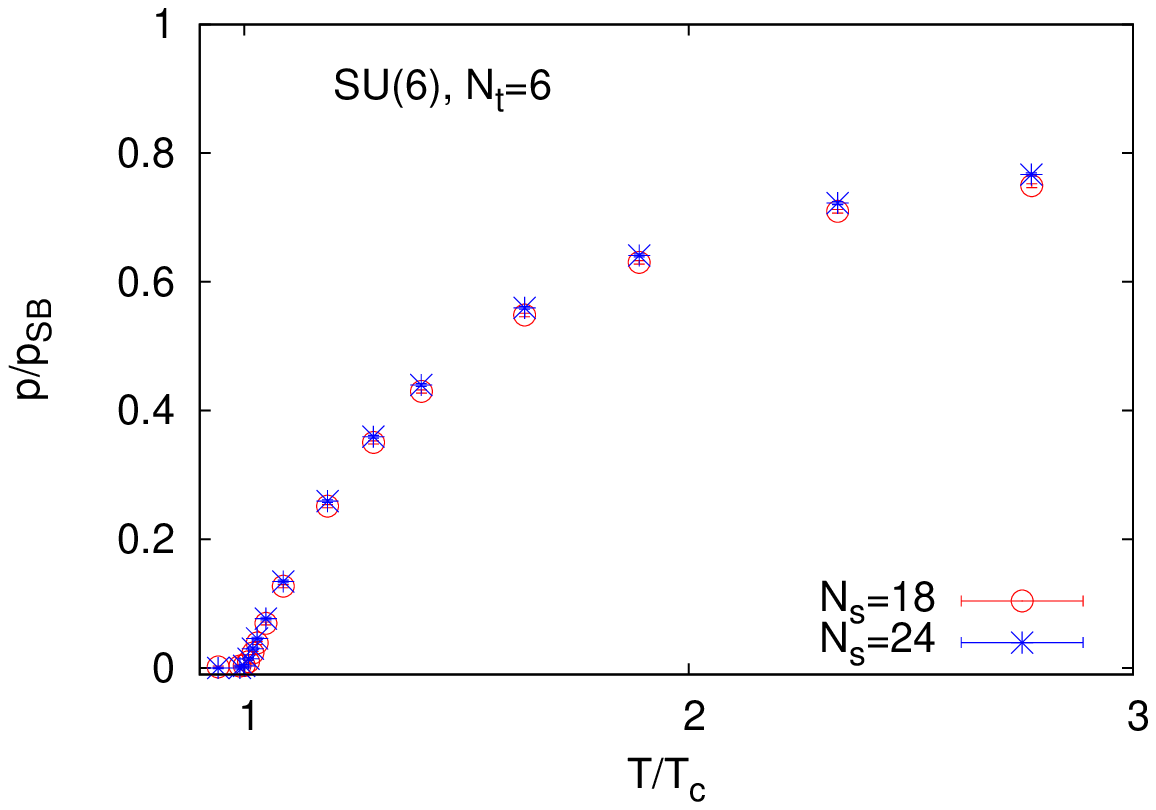}}
\scalebox{0.6}{\includegraphics{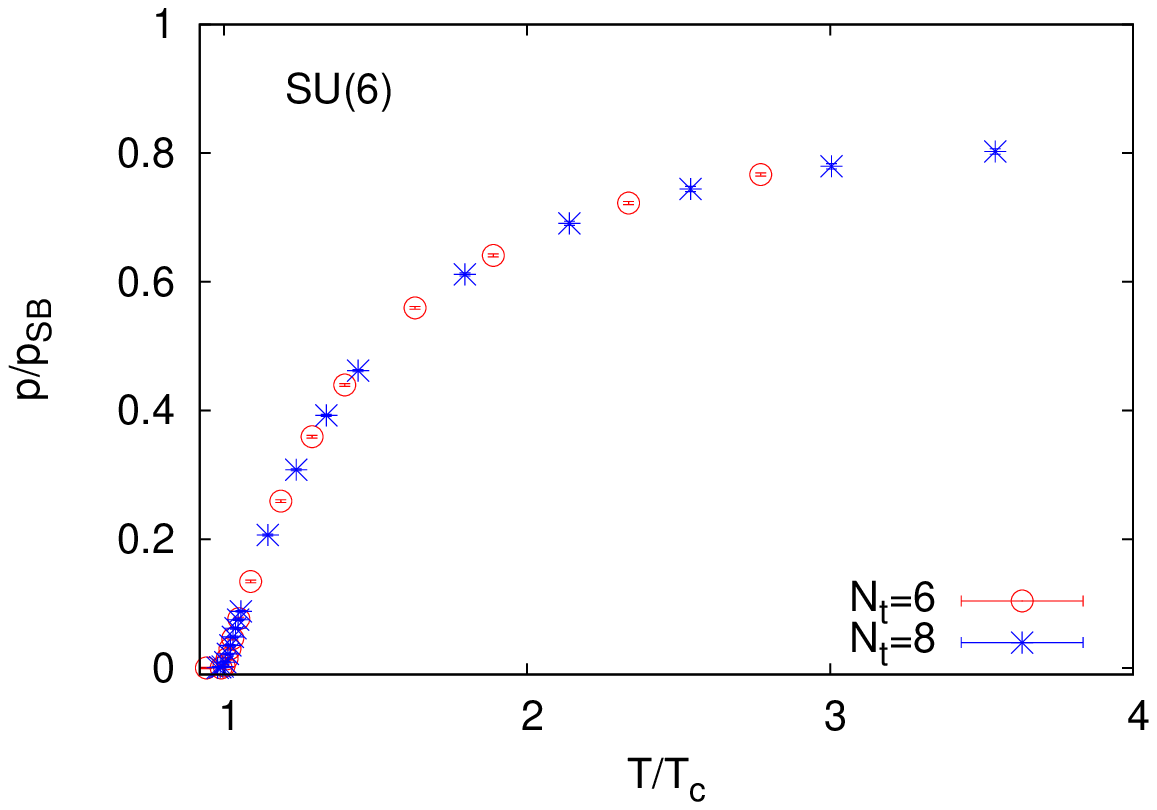}}
\end{center}
\caption{Study of systematics in the calculation of pressure for SU(4)
  and SU(6) gauge theory. Both finite size and finite lattice spacing
  effects seem to be under control.}
\label{fig.pressyst}
\end{figure}
 
\begin{figure}[bhpt]
\begin{center}
\scalebox{0.6}{\includegraphics{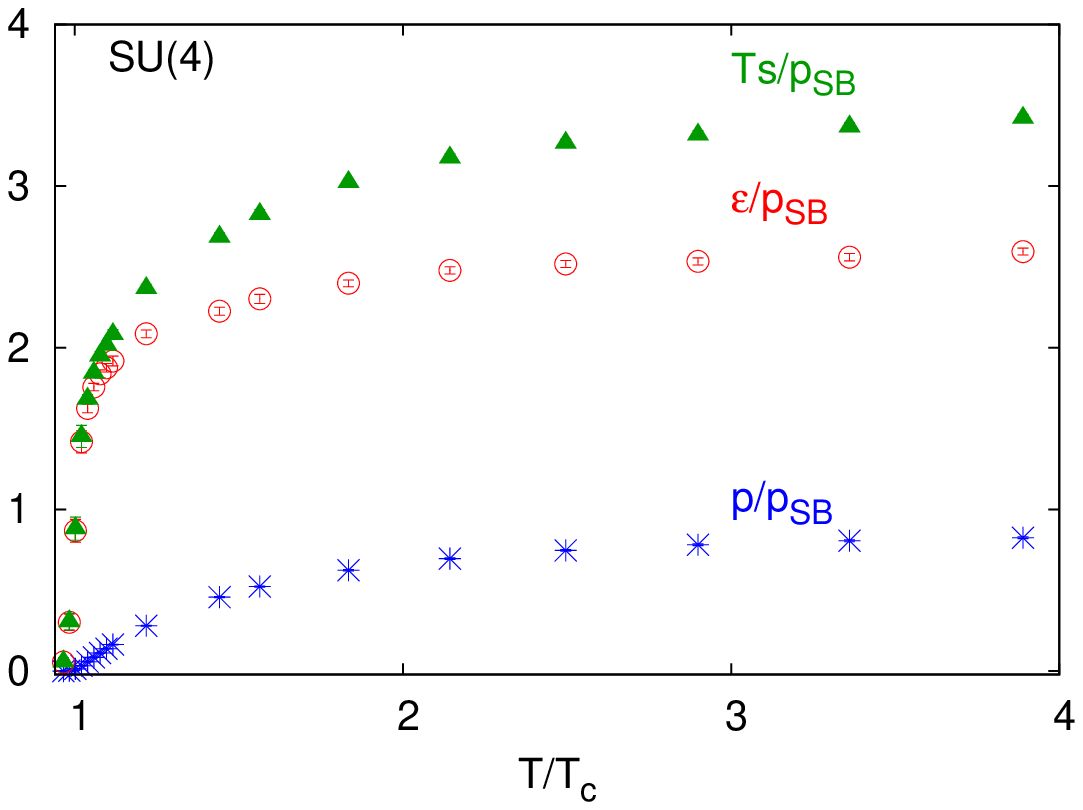}}
\scalebox{0.6}{\includegraphics{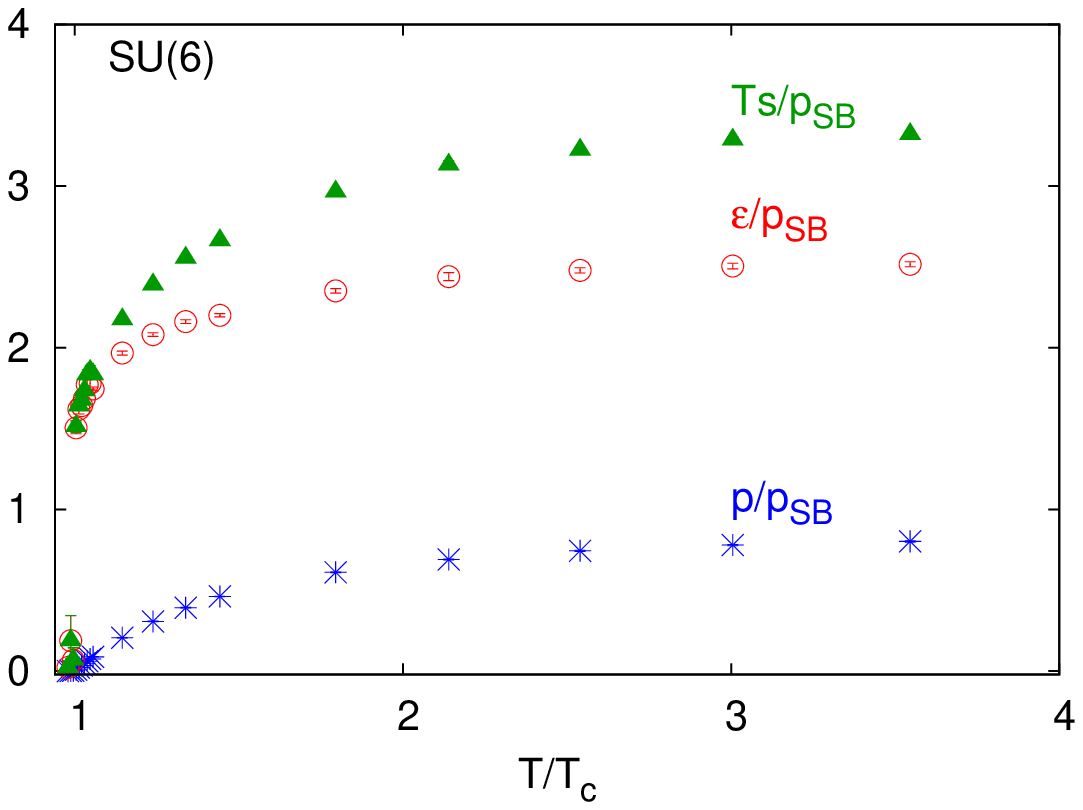}}
\end{center}
\caption{Results for thermodynamic quantities in SU(4) (left) and SU(6)
  (right) gauge theories. Shown are the pressure, energy density and
  entropy density, in units of $\psb$. The ideal gas limits for these
  quantities should be 1, 3 and 4 respectively.}
\label{fig.thermo}
\end{figure}

The pressure is calculated using the method outlined in (\ref{integral}).
The integration requires interpolation of the measured points, and there
could be a systematic error arising from this. We have estimated this
error by comparing linear and quadratic interpolations. We found that
point by point the error is small. Since the integration errors increase
over the range of $\beta$ and reach a maximum at the highest $\beta$
that we use, it is sufficient to report the magnitude of that error
compared to the statistical uncertainty. We can estimate the
significance of the error by the $t$-statistic---
\beq
   t = \frac{\left|I_2-I_1\right|}{\sqrt{\sigma^2(I_2)+\sigma^2(I_1)}}
\label{signif}\eeq
where $I_2$ is the integral estimates through a quadratic interpolation,
$I_1$ is the estimate using a linear interpolation and $\sigma^2(I_{1,2})$
is the statistical error in the estimate of $I_{1,2}$. This measure for SU(4)
is 1.03 and for SU(6) it is 0.007. This source of error is therefore almost
negligible. As a result, the systematic error is almost entirely due to
the neglect of $p_0$, the pressure at the lowest temperature where
the integration is started.

Figure \ref{fig.pressyst} shows the cutoff and volume dependence in the
calculation of pressure for the SU(4) and SU(6) theories.  The results
are normalized by the known (asymptotic) finite cutoff correction for
an ideal gluon gas, $\psb$, which was described earlier.  Hence the
pressure, so normalized, should go to unity. We find that finite volume
effects are negligible. Finite lattice spacing effects also turn out to
be negligible once we normalize the pressure by $\psb$
\footnote{In \cite{barak} the lattice spacing
dependence of $\psb$, \ie, $G(a)$ in eq.\ (\ref{lsb}), was instead
removed from the measured $p$.}. 
These results
indicate that it is safe to identify the continuum limit with the 
$N_t=8$ measurements.
Since the volume dependence is negligible,
at each $T$ we use the largest volume on which reliable results are
available as an indication of the thermodynamic limit.

At asymptotically large $T$ the $\sun$ gauge theory should go over to
the ideal gluon gas. However, even at the highest temperatures which we
have probed, \ie, $T\simeq4T_c$,
the ratio $p/\psb$ is far from unity (see Figure \ref{fig.pressyst}).
Given $p/T^4$ and $\Delta/T^4$ we obtain also the other bulk thermodynamic
quantities: $\epsilon/T^4$ and $s/T^3$. These results are collected
in Figure \ref{fig.thermo}. One sees clear deviations from the ideal gas
limit in these two quantities as well. All three quantities also show a
very slow rise throughout the measured range of $T$. Both $\epsilon$ and
$s$ show a rapid jump near $T_c$, stronger for SU(6) than for SU(4).

\section{$N_c$ scaling of thermodynamic observables}
\label{sec.nc}

\begin{figure}[t]
\begin{center}
\scalebox{0.6}{\includegraphics{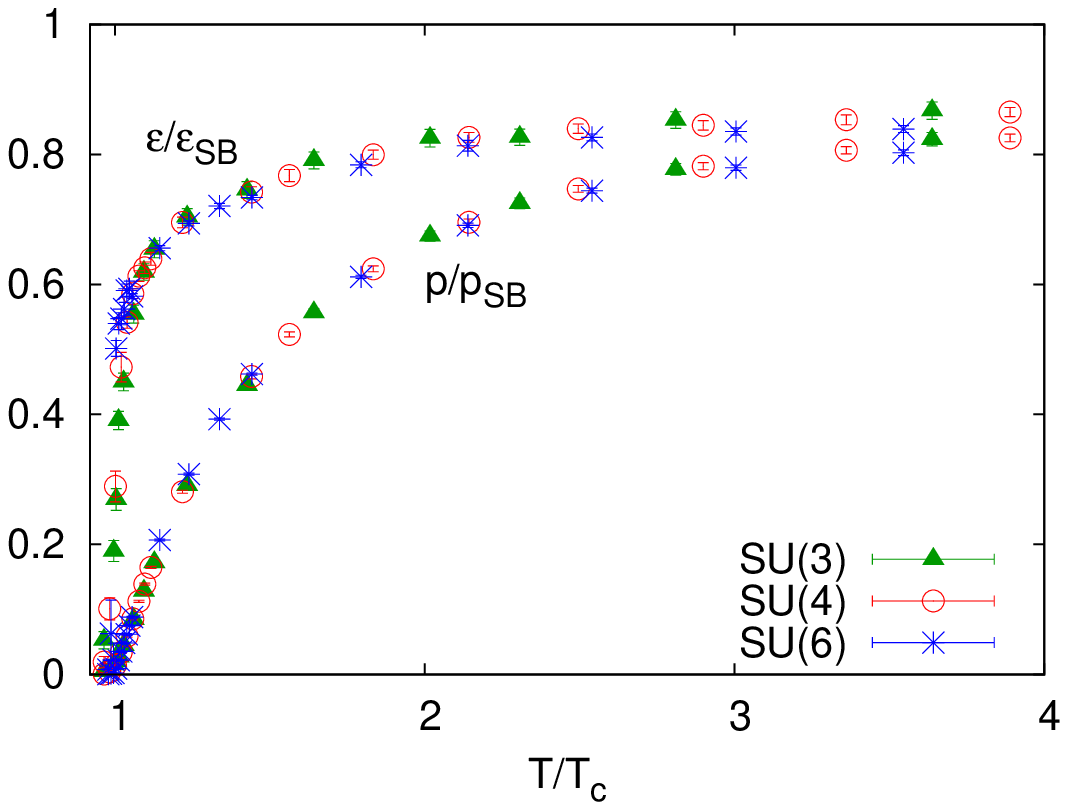}}
\scalebox{0.6}{\includegraphics{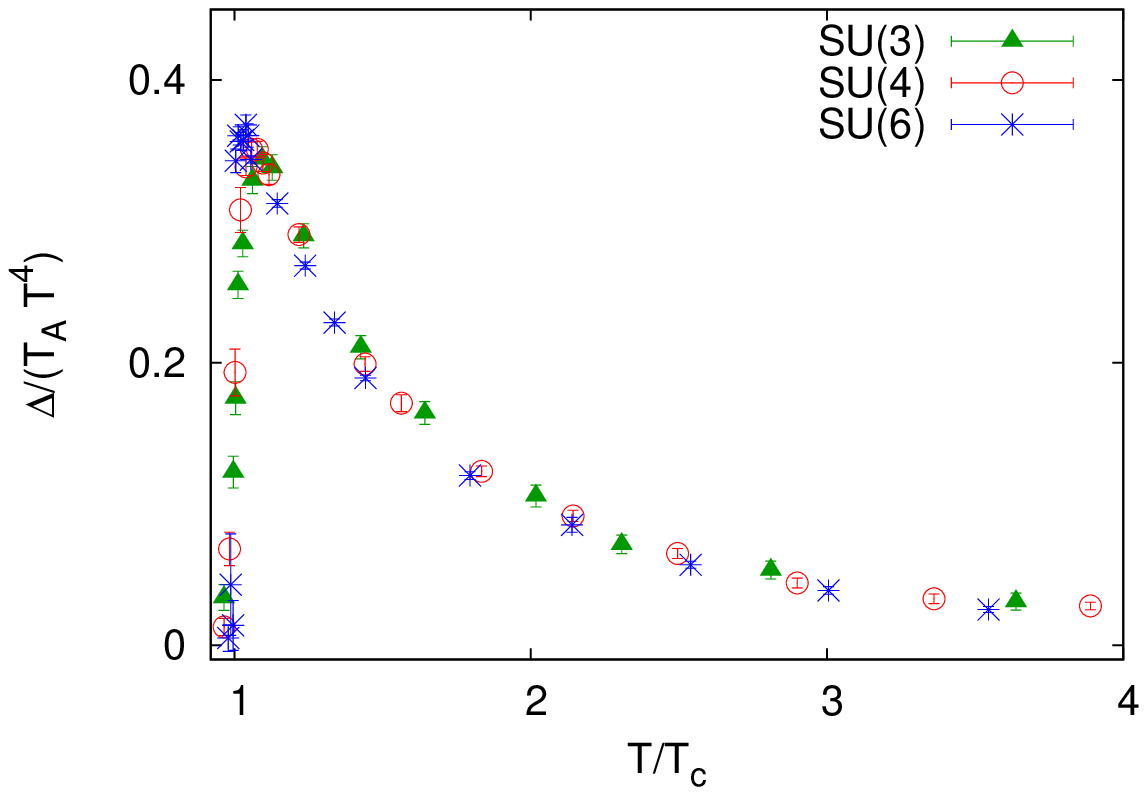}}
\end{center}
\caption{The first panel shows the energy density and pressure in $\sun$
  pure gauge theories, normalized to the corresponding ideal gas values.
  The second panel shows $\ept$, normalized by $d_A=N_c^2-1$, for the
  same theories.}
\label{fig.suNthermo}\end{figure}

\begin{figure}[thpb]
\begin{center}
\scalebox{0.9}{\includegraphics{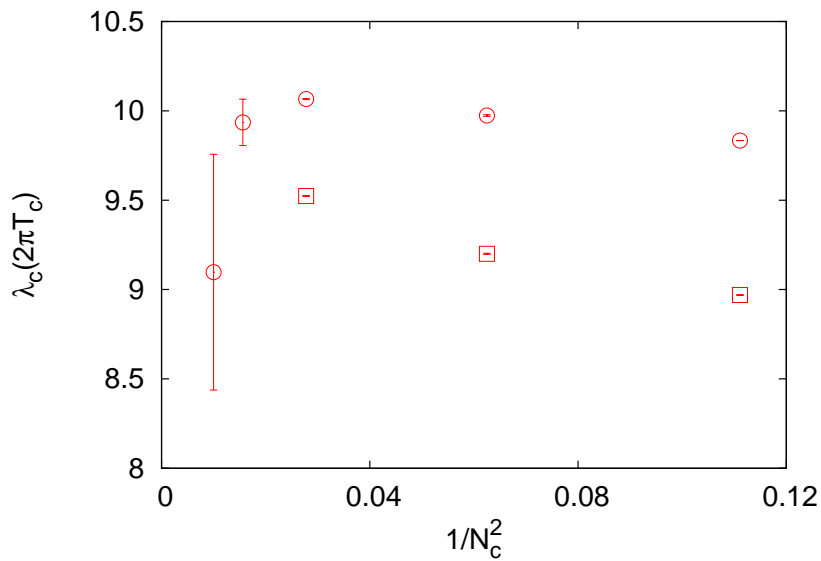}}
\end{center}
\caption{The value of the 't Hooft coupling at the finite temperature
   transition, $\lambda_c$, for different $N_c$. The renormalized gauge
   coupling is evaluated at scale of $2\pi T_c$, so that the result is
   independent of the lattice spacing. The boxes denote results obtained
   using the non-perturbative beta-function and the circles using the
   two-loop beta-function in the V-scheme.}
\label{fig.thooftcrit}\end{figure}

\begin{figure}[bhpt]
\begin{center}
\scalebox{0.6}{\includegraphics{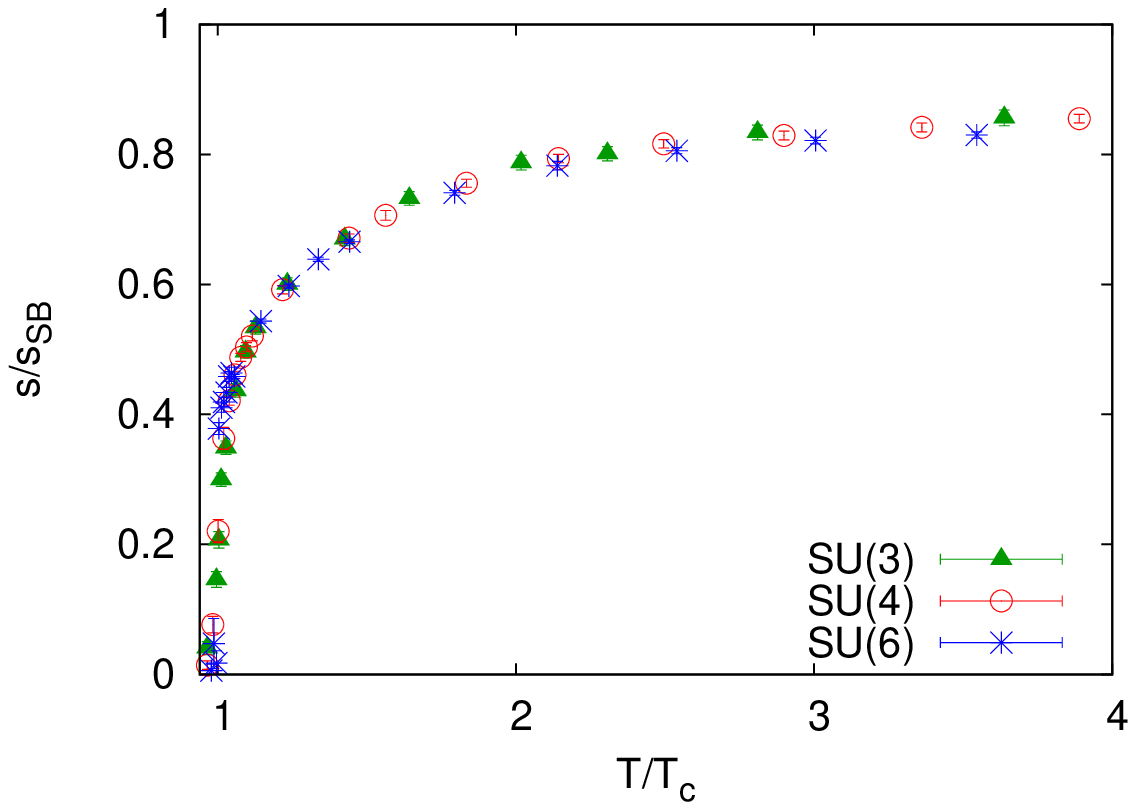}}
\scalebox{0.6}{\includegraphics{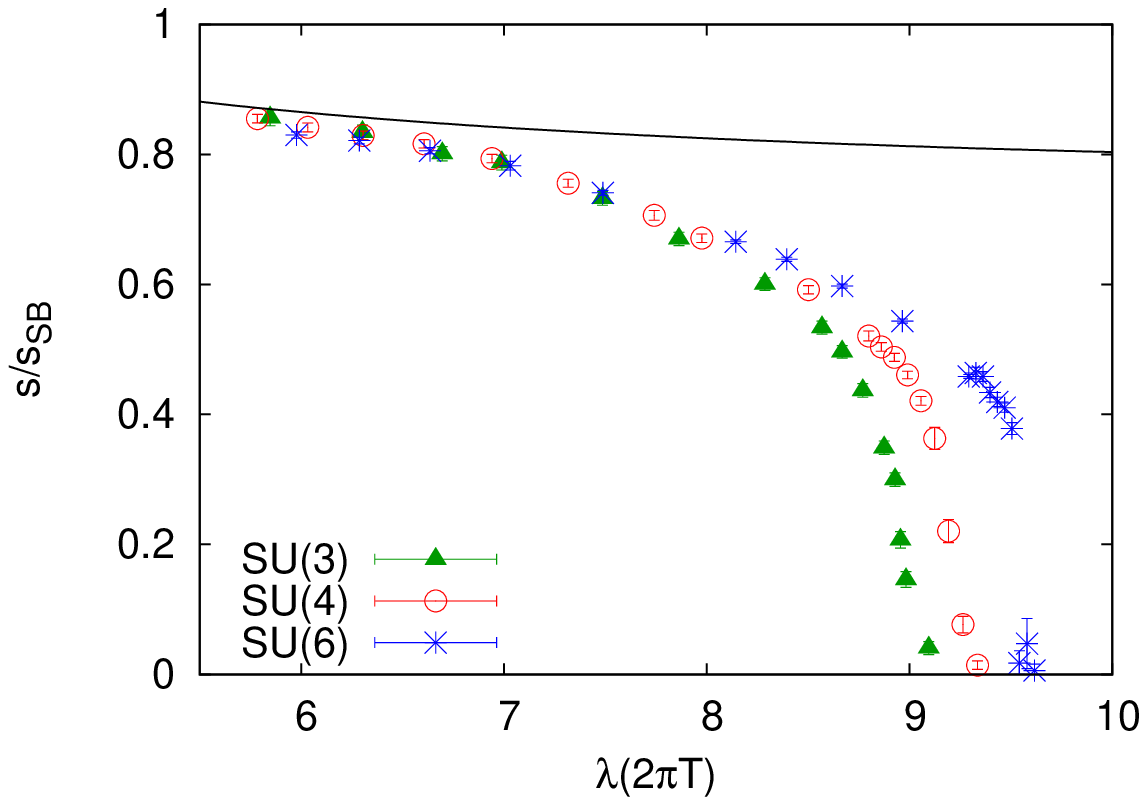}}
\end{center}
\caption{The scaled entropy density, $s/\ssb$, as a function of $T/T_c$
   and the `t Hooft coupling $\lambda=N_c g_R^2$. The gauge coupling is
   evaluated at the scale $2\pi T$ using the non-perturbative beta-function
   for the appropriate theory.}
\label{fig.thooft}\end{figure}

As discussed earlier, we distinguish between strong $N_c$ scaling and
't Hooft scaling. The first is scaling with $N_c$ of thermodynamic
quantities at fixed $T/T_c$, and the second, the scaling with $N_c$ at
fixed $\lambda$.  We examine scaling by combining our results for the
continuum limit of bulk thermodynamic quantities for $N_c=4$ and 6, with
a reanalysis of the older $N_c=3$ data of \cite{boyd} using our techniques.

Strong $N_c$ scaling has been observed on the lattice in many
contexts \cite{teper1,teper2,barak,panero}.  The continuum limit
of bulk thermodynamic quantities that we have extracted are also
consistent with this limiting procedure, except for $N_c=3$ near $T_c$.
In Figure \ref{fig.suNthermo} we show the energy density and pressure,
each normalized by its ideal gas value, for $N_c=3$, 4 and 6. As noted
earlier, there are clear deviations from the ideal gas behaviour, but
the scaled quantities are almost independent of $N_c$. Since the ideal
gas values scale with $d_A$, one expects that at large $T$ all bulk
thermodynamic quantities should scale accordingly.

We see that $p/\psb$ is independent of $N_c$ over the whole temperature
range above $T_c$ within the accuracy of our measurement. Such a statement
is also true for the energy density except when $T$ is close to $T_c$.  Close to $T_c$
the energy density does not scale as the ideal gas, \ie, as $d_A$, between
$N_c=3$ and larger values of $N_c$.  This is, of course, a consequence of
the fact that the latent heat density does not scale as $d_A$ for $N_c=3$
(see Table \ref{tbl.lht}).  Consequently, $\Delta/T^4$ also fails to
scale with $d_A$ in the vicinity of $T_c$.  In Figure \ref{fig.suNthermo} we
also show that the peak of $\Delta/(T^4d_A)$ is smaller for $N_c=3$ than for
$N_c=4$ and 6 (the latter two are almost identical in value).  We also
see that the peak is rounded and shifted away from $T_c$ at $N_c=3$.
As discussed in Section \ref{sec.lht}, this could be due to finite-size
effects, since $N_c=3$ has a weaker first-order phase transition. With
this exception, strong $N_c$ scaling seems to work very well for bulk
thermodynamic quantities; the sub-leading corrections are too small to
be seen over the statistical errors.

Next we turn to evidence for 't Hooft scaling. Since the critical
point is known very precisely for several $N_c$ \cite{scaling}, we test
this scaling using $\lambda_c=N_cg_R^2(2\pi T_c)$ \footnote{A similar
test was performed with the bare coupling in \cite{scaling}. For the
running coupling $2\pi T_c$ is a nearly optimal scale \cite{schroder},
and data for different $N_t$ collapse to a single value with this choice
of scale. The exact value of $\lambda_c$ is, of course, dependent on
the scheme and the precise choice of scale.}. The change of $\lambda_c$
with $N_c$ is very much larger than the statistical errors, and is seen
using both the non-perturbative and the two-loop beta-functions, although
it is somewhat larger with the former. The change is non-monotonic when
the two-loop beta-function is used.  In this context, we recall the
result shown in the appendix: that the non-perturbative beta-function is
preferred near $T_c$.  Restricting ourselves to using only this leaves
the three smallest values of $N_c$.

Even with these three, it is clear from Figure
\ref{fig.thooftcrit} that the data do not fall on a straight line and
hence a single correction term does not suffice. Using a second correction
term, a description of the confinement-deconfinement phase boundary,
\ie, the variation of $\lambda_c$ with $1/N_c^2$, is---
\beq
   \lambda_c = \begin{cases}
      9.8771 (4) - \frac{14.2562 (2)}{N_c^2} + \frac{54.7830 (2)}{N_c^4}
        & \text{(non-perturbative),} \\
      9.9904 (6) + \frac{1.2081 (3)}{N_c^2} - \frac{23.5709 (3)}{N_c^4}
        & \text{(two-loop).}
    \end{cases}
\label{phasebdry}\eeq
Since the fit is linear in the parameters, the formal solution for
$\chi^2$ minimization can be written down along with parameter errors.
At $N_c=3$ the values of the ${\cal O}(1/N_c^2)$ and ${\cal O}(1/N_c^4)$
terms are comparable. Although they are small corrections to the leading,
${\cal O} (N_c^0)$ term, this behaviour of the series could indicate
that $N_c=3$ lies near the radius of convergence of the series around
$N_c=\infty$.

If this is so, then summing three terms of the series is not numerically
accurate. However, with three pieces of data fitting a large number of
terms is an ill-conditioned problem. As a result, one would do better
to fit a resummation of the series, provided such a resummation has a
small number of parameters. Unfortunately, there is no theory for the
shape of the phase boundary. In its absence we try the usual trick of
estimating a Pad\'e resummation of the series. With three terms the best
that we can try to do is to fit the lowest order expansion
\beq
   \lambda_c(N_c) = \lambda_c(\infty) + \frac{a/N_c^2}{1+N_*^2/N_c^2}.
\label{pade}\eeq
The best fit gives $N_*\simeq4$, roughly consistent with the series
analysis. As a result, the fitted value of $\lambda_c(\infty)$ is
sensitive to the form of the remaining function, and cannot be reliably
extracted using data for $N_c$ near 4.  It would be useful to improve
the computations with $N_c>6$ in order to extract this quantity with
better accuracy.

Next, we extend such a test to a bulk thermodynamic quantity.  Our results
for $s/\ssb$ as functions of $T/T_c$ and $\lambda(2\pi T)$ are shown in
Figure \ref{fig.thooft}.  As we show in the figure, strong scaling holds
with good precision since $s/\ssb$ is almost independent of $N_c$ down
to the smallest temperatures that we have studied.  For scaling at fixed
$\lambda$, convergence of the series is clearly bad for $\lambda\simeq9$
or larger, because the theories for different $N_c$ begin to drop out
of the plasma phase; figure \ref{fig.thooftcrit} shows that theories with
smaller $N_c$ drop to the confined phase at smaller $\lambda$. At any
fixed $\lambda$ in this region one has to go to $N_c$ large enough that
the theories are all in the same phase before one can observe good
scaling with $N_c$. We also find that the convergence of the series in
$1/N_c^2$ is acceptable when $\lambda<8.6$ \footnote{This corresponds
to $T>1.1 T_c$ for SU(3) and $T>1.25 T_c$ for SU(6). For strong scaling,
in contrast, the problematic region lies in the vicinity of $T_c$ in an
interval which shrinks as $N_c$ grows.}. In the range $8.6<\lambda<9$,
the physically interesting theory with $N_c=3$ is close to the radius
of convergence of the series expansion, and the effect of the correction
terms is large.

Figure \ref{fig.thooft} also displays a comparison of the scaled entropy
with predictions in a ${\cal N}=4$ supersymmetric Yang-Mills theory, 
computed \cite{gubser} using the AdS/CFT correspondence---
\beq
   \frac s{\ssb} = \frac34 
      + \frac{45}{32}\zeta(3)\left(\frac1\lambda\right)^{3/2} + \cdots.
\label{sym}\eeq
Although one does not expect this computation to be valid in the realistic
non-supersymmetric theories under investigation, it is sometimes said
to agree with lattice results.  Here we show that the agreement is
poor, except at the highest possible temperatures. It has been argued
\cite{panero} that more realistic AdS-QCD models should be used for such
a comparison; the analysis of such models lies beyond the scope of this
paper.

\section{Summary and Discussion}
\label{sec.dis}

\begin{figure}
\begin{center}
\scalebox{0.9}{\includegraphics{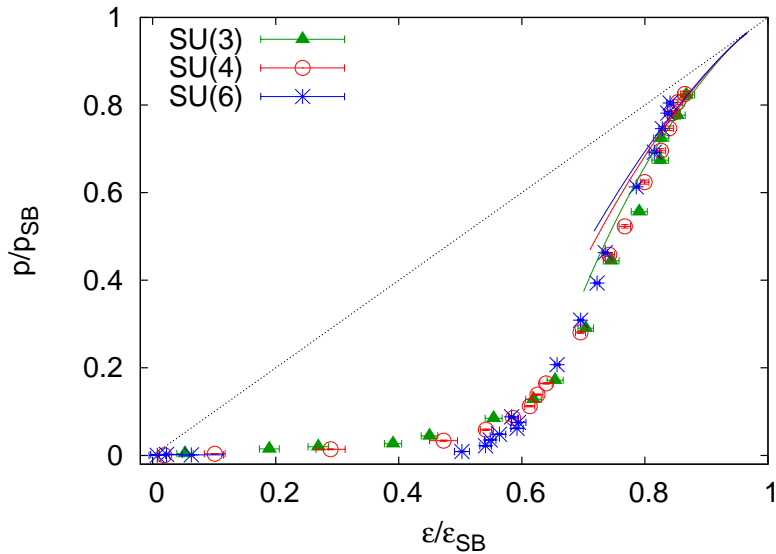}}
\end{center}
\caption{Approach to conformality in $\sun$ gauge theories. The diagonal
  line contains results from all possible conformal theories; the upper
  right end is the special case of a free massless theory. Also shown
  are the weak coupling results from \cite{mikko} (with increasing $N_c$
  the curves move up monotonically). Since the lattice
  data lie closer to the weak-coupling results than to the diagonal
  line, there seems to be no window where a strongly coupled conformal
  theory describes $\sun$ pure gauge thermodynamics.}
\label{fig.conform}
\end{figure}

In this paper we have studied the thermodynamics of the \plasma, by
numerical simulations of SU(4) and SU(6) gauge theories, and comparing
them with a reanalysis of existing data \cite{boyd} for SU(3). Our focus
is on taking the continuum and thermodynamic limits, by using multiple
spatial volumes at each cutoff and by using significantly smaller
cutoffs than used previously for the EOS of $N_c>3$.  As discussed in
the introduction, in terms of statistics, lattice spacing and spatial
volumes, this work brings the study of the thermodynamics of pure gauge
$\sun$ theories with small $N_c>3$ at par with the state of the art for
$N_c=3$, while also throwing some new light on the SU(3) theory.

We had shown in an earlier study of the deconfinement transition
\cite{scaling} that observations made with finite lattice spacing could be
continued to the continuum using the renormalization group equations. When
measurements are made with lattice spacing $a\le1/(8T_c)$ we had found
that the two-loop beta-function suffices. When the lattice spacing
$a=1/(6T_c)$ is used a non-perturbative beta-function was introduced
which could be used to continue the
lattice results to the continuum limit \cite{scaling}. In this paper we
used these earlier results to obtain the continuum thermodynamics of pure
gauge SU(4) and SU(6) theories.  We also made a reanalysis of the older
SU(3) data using this technique (as detailed in the appendix).

One of our important results (see Section \ref{sec.lht}) is an extraction of
the latent heat of the deconfinement transition,
$\latent/T_c^4$, for $N_c=3$, 4 and 6. We found that $\latent/(d_AT_c^4)$,
where $d_A=N_c^2-1$,
increases between $N_c=3$ and larger $N_c$, indicating that the
first-order transition grows stronger with increasing $N_c$. Our results are
compatible with \cite{teper2}. One expects
stronger transitions to have smaller finite volume effects, and our
observations support this notion. We found some scale breaking
in the measurement of $\latent/T_c^4$, and observed that $\latent/\epm$
shows better scaling properties. We also saw that in the large $N_c$ limit
one has $\latent/(d_AT_c^4)=0.388\pm0.003$ (see eq.\ \ref{fitted}).

Further study of bulk thermodynamic quantities started with measurements
of $\ept$ (details are given in Section \ref{sec.e3p}). We found
good scaling of this quantity, and a reliable continuum limit, for
$T>T_c$. For SU(3) some scale breaking is observed very close to $T_c$ where the
peak of this quantity lies. The cause remains obscure, although there
are some indications that lead us to conjecture that this could be due
to finite volume effects. Future studies are planned to understand this
remaining ambiguity. In a range of temperature up to $4T_c$ we found that
$\Delta\propto T^2$ \cite{ogilvie,pisarski}, or, possibly, slower.

The pressure, $p/T^4$, was obtained using the so-called integral method
(see Section \ref{sec.eos}). Finite volume and lattice spacing effects
in this measurement are under good control. We extracted the energy
density, $\epsilon/T^4$, and the entropy density, $s/T^3$, using these
two primary measurements. In the whole range of $T$ all these quantities
lie substantially below the ideal gas values (see Figures \ref{fig.thermo}
and \ref{fig.suNthermo}).  Nevertheless, $p/T^4$, considered as a function
of $T/T_c$, scales very well with $d_A$ \cite{barak,panero}. Sub-leading
corrections in $N_c$ are hard to see at the level of accuracy we have
reached (see Figure \ref{fig.suNthermo}). Similar scaling with $d_A$
is also seen for $\epsilon/T^4$ and $s/T^4$, except, possibly, in a
small region near $T_c$.

In Figure \ref{fig.conform} we present a plot of the normalized energy
density, $\epsilon/\esb$ against the normalized pressure, $p/\psb$
following \cite{swagato}. The diagonal line is the line of all conformal
theories, with the ideal gluon gas being one special point on it. Weak
coupling results \cite{mikko} lie below this. One sees that the lattice
data for $N_c=3$, 4 and 6 lie further below. Indeed the topology of
these relations is such that the weak coupling predictions are always a
better approximation to the lattice data than conformal theories. This
figure gives clear evidence that there is no window of temperature in
which the EOS can be described by a strongly-coupled conformal theory
better than by weak coupling theory.

Yet another reason for making high-precision measurements of bulk
thermodynamics for $N_c\ne3$ is to understand the usefulness of the
large-$N_c$ limit (see Section \ref{sec.nc}). In agreement with previous
results, we find that the strong scaling limit obtained by taking
fixed $T$ and $N_c\to\infty$ works very well. Corrections in powers of
$1/N_c$ are small, as a result of which the large-$N_c$ results for the
entropy density, for example, can
be directly applied for $N_c=3$ with about 1\% error. However, when the
same data is analyzed as a function of the 't Hooft coupling, the finite
$N_c$ corrections are large for $\lambda>8.6$. The phase boundary for the
large $N_c$ theory, expanded in powers of $1/N_c^2$, seems to have a
radius of convergence smaller than 1/9 (see eq.\ \ref{pade}).
Since this observation has ramifications
for all models of thermal QCD which proceed from the large-$N_c$ approximation,
including string-based models, we plan further measurements in the near
future to explore the applicability of the 't Hooft limit.

We thank Mikko Laine for providing us with the weak coupling results of
Fig. \ref{fig.conform} and Rob Pisarski for comments. The computations
were carried out on the workstation farm of the department of theoretical
physics, TIFR and the Cray X1 of the ILGTI. We thank Ajay Salve for
technical support.

\appendix
\section{The beta-function}
\label{sec.beta}

\begin{figure}[tpb]
\begin{center}
\scalebox{0.6}{\includegraphics{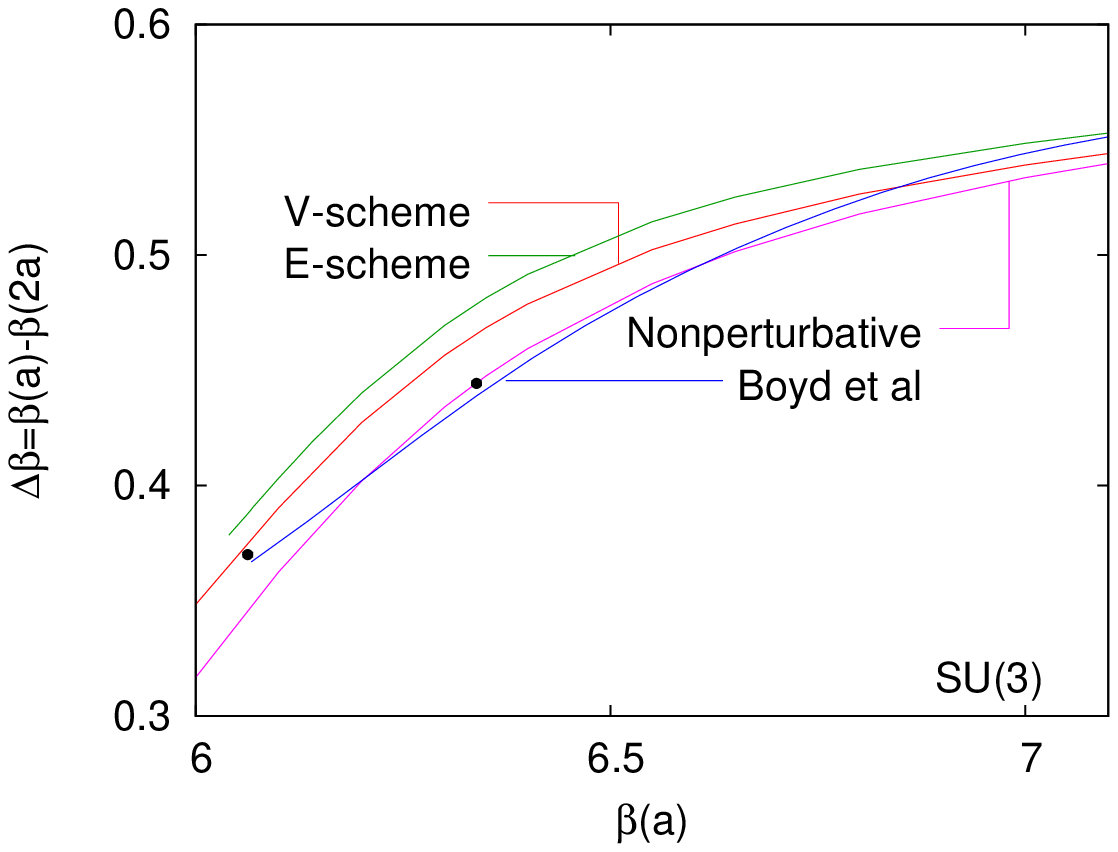}}
\scalebox{0.6}{\includegraphics{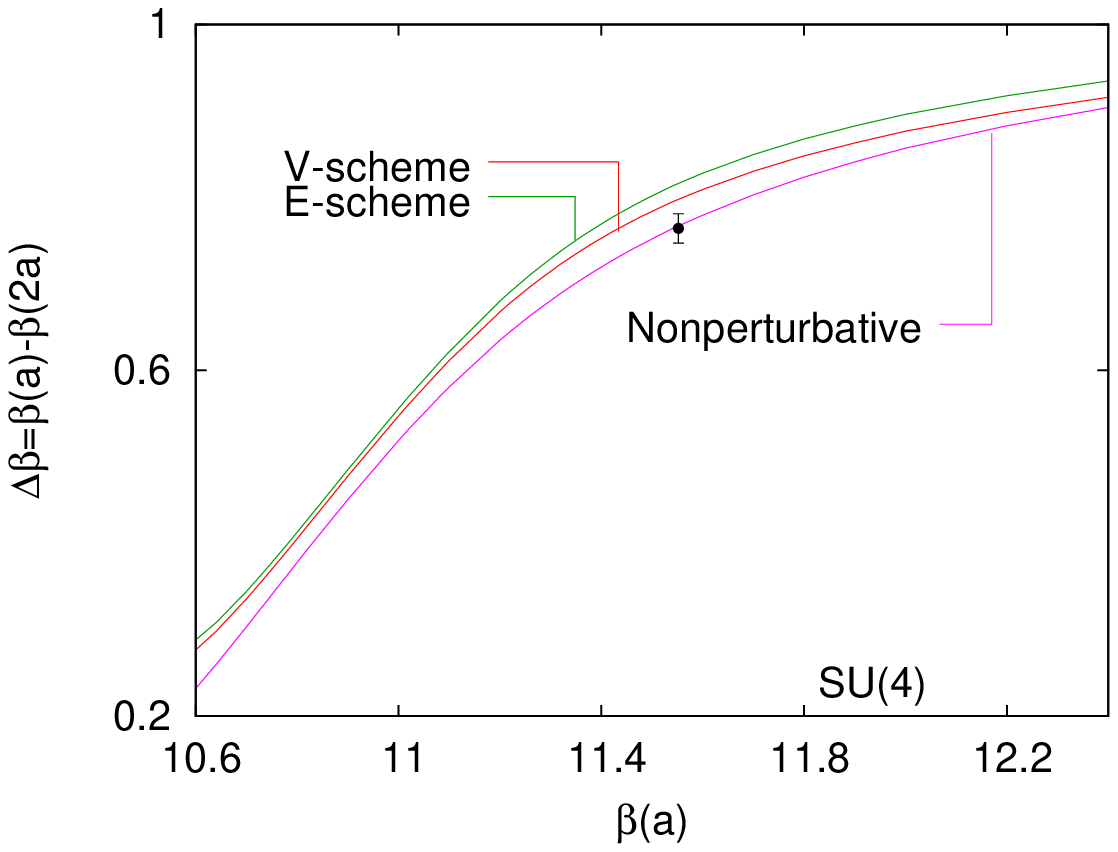}}
\end{center}
\caption{The step scaling function in SU(3) (left) and SU(4) (right)
  gauge theory. The data points are for the location of the deconfinement
  transition. For SU(3) they are taken from \cite{boyd,fingberg} and for
  SU(4) from \cite{scaling}. The non-perturbative beta-function of
  \cite{boyd} is tuned for use with $a\le a_c(4)$. We have used a
  non-perturbative beta-function \cite{precise,scaling} which can
  be used for $a<a_c(6)$.}
\label{fig.stepscl}
\end{figure}

\begin{figure}[bpt]
\begin{center}
\scalebox{0.6}{\includegraphics{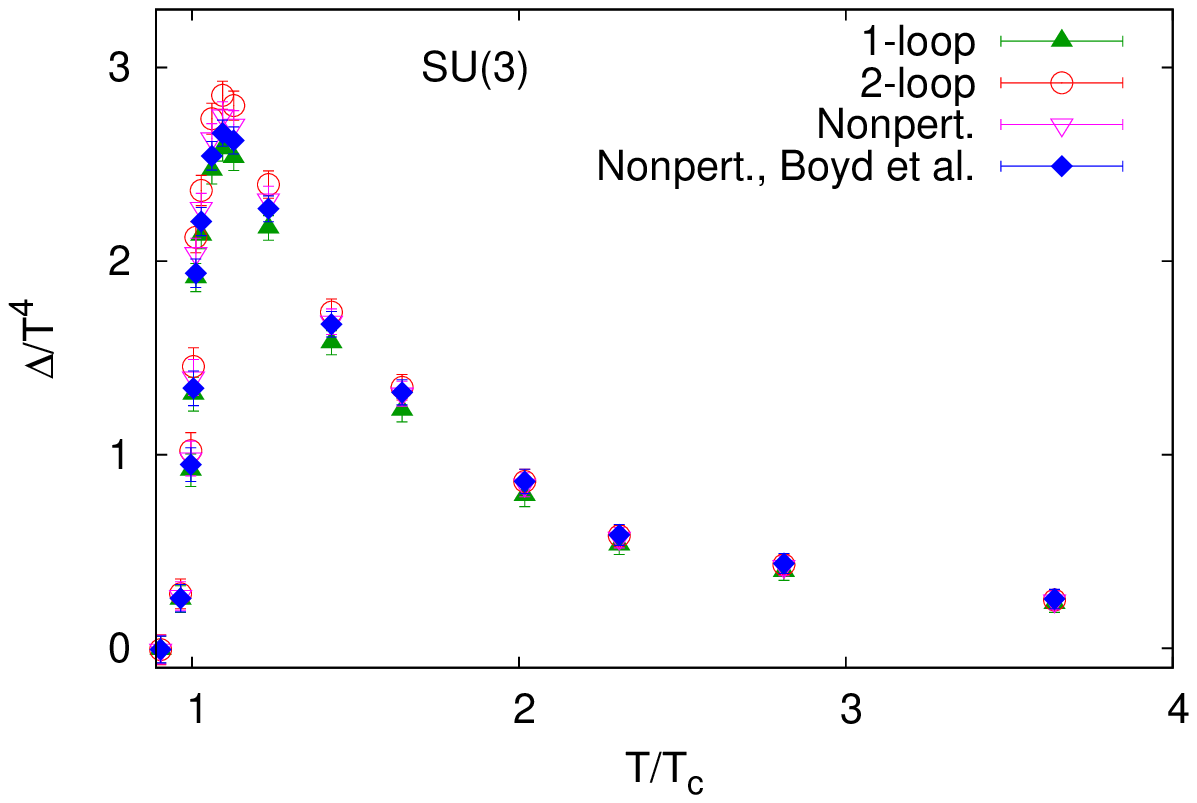}}
\scalebox{0.6}{\includegraphics{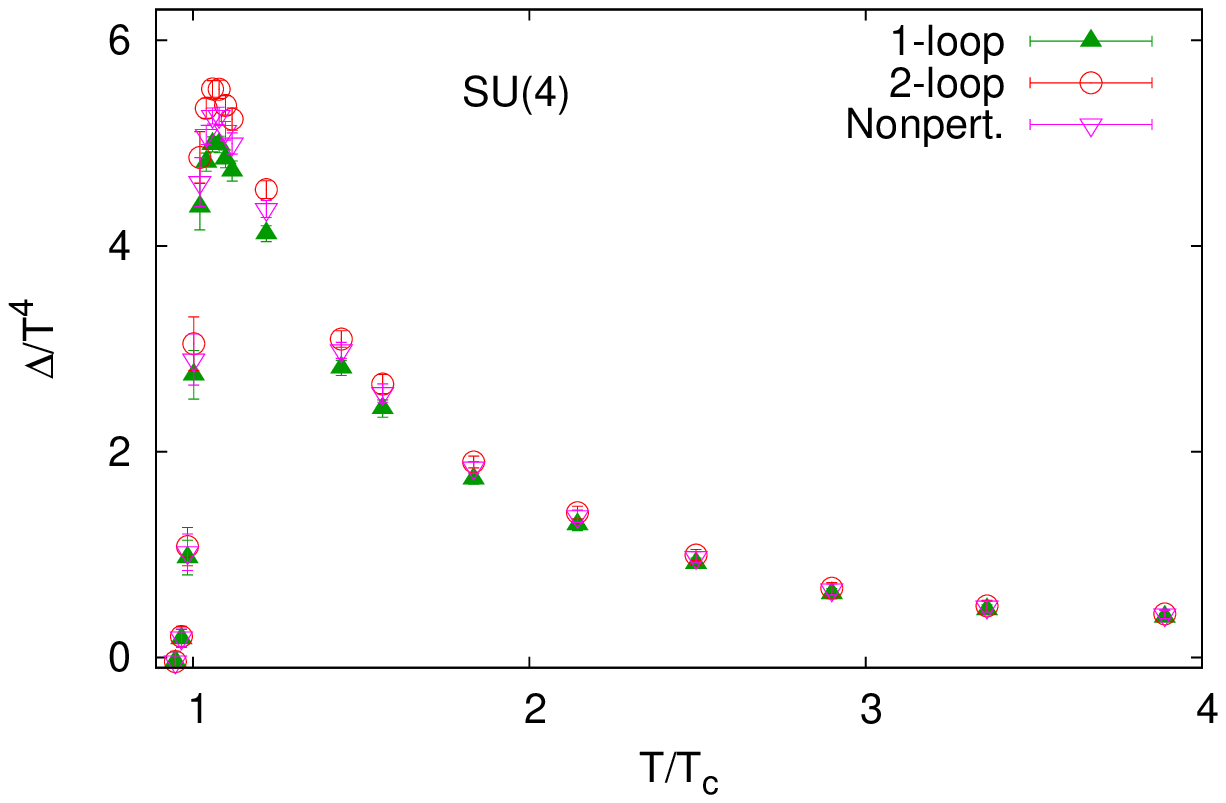}}
\end{center}
\caption{$\ept$ for SU(3) gauge theory on $32^3\times8$ lattices (left) and
SU(4) on $30^3\times8$ lattices (right). Shown are the analyses using various
different beta-functions (based on the V-scheme). For SU(3), where the data
is taken from \cite{boyd} we also show the analysis using the non-perturbative
beta-function of that paper.}
\label{fig.betafndep}
\end{figure}

The lattice theory is cut off at a length scale of $a$. When $a$ is small
various quantities have a perturbation expansion in $g_R$, the renormalized
coupling determined at a scale $\kappa a$ (where $\kappa$ can depend on the
scheme). Then the derivative in eq.\ (\ref{e3p}) can be written as
\beq
   \frac{\partial\beta}{\partial\ln a} =
   \frac{\partial\beta}{\partial g_R} \frac{\partial g_R}{\partial\ln a},
\label{defbetafn}\eeq
where the last factor is the negative of the beta-function.
Due to the fact that $\sun$
gauge theories are asymptotically free, one expects that a weak-coupling
determination of the beta-function should suffice when $a$ is small
enough.  In \cite{scaling} it was shown that the two-loop beta-function is
a sufficient description of the flow of the renormalized coupling at the
scale of $a\le1/(8T_c)$.  Confidence in the efficacy of two-loop scaling
is enhanced by the fact that the scheme dependence in the extraction of
the QCD scale was small.

Although the two-loop beta-function was insufficient to describe the flow
of the coupling at larger $a$, it was shown that for $a\simeq a_c(6)$,
where $a_c(N_t)=1/(N_t T_c)$, a simple correction of the form
\beq
   a_c(N_t)\Lambda = R\left(\frac1{\beta_0g_R^2(a_c)}\right)\,
    \left[1+\frac{c_2}{N_t^2}\right],\qquad{\rm with}\qquad
   R(x) = {\rm e}^{-x/2} x^{\beta_1/2\beta_0^2}
\label{betafn}\eeq
suffices, where the two-loop beta-function is
$-\beta_0g_R^3-\beta_1g_R^5$ (for an alternative approach see \cite{alphas}).  The values of $c_2$ were presented in
Table (\Rmnum{4}) of \cite{scaling}.
In this paper the integration of the beta-function
is started from the scale $a_c(8)$.  In the calculation of section
\ref{sec.e3p}, since we have taken data at lattice spacings as low as
$a_c(6)$, we have used such a non-perturbatively corrected beta-function
in the V-scheme
(this differs from the conventions of \cite{qm09}).  
Any non-perturbative beta-function will include finite
lattice spacing corrections \cite{allton} into the scaling, just as the
above function does. Such corrections are non universal.

A simplified version of the tests of scaling in \cite{precise,scaling} can
be presented using the step-scaling function $\Delta\beta(\beta)$. This
is the change in the bare coupling, $\Delta\beta$, required to reproduce
the physics observed at a bare coupling $\beta$, when the lattice
spacing is doubled. If $\beta$ is chosen to be the lattice coupling
where the deconfinement transition is observed for a given $N_t$,
then $\beta-\Delta\beta$ is the lattice coupling at the deconfinement
transition when $N_t$ is changed to $N_t/2$. In Figure \ref{fig.stepscl}
we show the result of using the beta-functions given above and the step
scaling function given in \cite{boyd}.

We also examined the sensitivity of the equation of state to the choice of the
beta-function.  Figure \ref{fig.betafndep} displays results for $\ept$
obtained in SU(3) and SU(4) gauge theory, for lattices with $a=1/(8T)$.
For the SU(4) theory, differences between the E-scheme and
the V-scheme are statistically insignificant at all temperatures.
Similarly, the difference between
these and the non-perturbative beta-function of (\ref{betafn}) are also
insignificant at these temperatures. 

The results are similar for the SU(3) gauge theory, where we have
re-analyzed the data of \cite{boyd}. The one-loop and the two-loop
beta-functions give coincident results for $T\ge1.5T_c$. In \cite{precise}
a non-perturbative beta-function of the form in (\ref{betafn})) was used
to describe scaling of $T_c$. The results of using this for $\ept$ are
shown. These three are close to each other, as is the result of \cite{boyd}.

\end{document}